\documentstyle[11pt,aaspp4]{article}
\font\sml=cmr10
\def\lsim{\lower.5ex\hbox{$\; \buildrel < \over \sim \;$}}
\def\gsim{\lower.5ex\hbox{$\; \buildrel > \over \sim \;$}}

\begin{document}
\tighten

\title{X-Rays from Accelerated Ion Interactions}

\author{Vincent Tatischeff}
\affil{Laboratory for High Energy Astrophysics\\ Goddard Space 
Flight Center, Greenbelt, MD 20771}

\author{Reuven Ramaty}
\affil{Laboratory for High Energy Astrophysics\\ Goddard Space 
Flight Center, Greenbelt, MD 20771}

\and

\author{Benzion Kozlovsky}
\affil{Sackler Faculty of Exact Science, Tel Aviv University\\ 
Ramat Aviv, Tel Aviv, Israel}

\begin{abstract}

We have developed in detail the theory of X-ray line and continuum 
production due to atomic interactions of accelerated ions, 
incorporating in our calculations information from a broad range of 
laboratory measurements. We applied our calculations to the Orion 
region from which nuclear gamma-ray lines were observed with the 
COMPTEL instrument on {\it CGRO}. The accelerated particles which 
produce this gamma-ray emission via nuclear reactions also produce 
X-ray lines via atomic interactions. We predict strong line emission 
in the range from 0.5 to 1 keV, mainly due to de-excitations in fast 
O ions. While much of the diffuse X-ray emission observed with ROSAT 
from Orion could be due to accelerated ions, the current X-ray data 
do not provide unambiguous signatures for such an origin. If future 
observations with high spectral resolution would confirm the 
predicted X-rays, the combined analysis of the X-ray and gamma-ray 
data will set important constraints on the origin of the accelerated 
particles and their interaction model. 

\end{abstract}


\eject

\section{Introduction}

Strong gamma-ray emission in the 3-7 MeV range has been detected 
from the Orion molecular cloud complex with the COMPTEL instrument 
on the {\it Compton Gamma Ray Observatory (CGRO}; Bloemen et al. 
1994, 1997). As the observed spectrum exhibits characteristic 
structures (Bloemen et al. 1997), this emission is most likely due 
to the superposition of nuclear gamma-ray lines, mainly the 4.44 MeV 
line from $^{12}$C and the 6.13, 6.92 and 7.12 MeV lines from 
$^{16}$O. Such line emission can only be produced by accelerated 
particle interactions. Gamma-ray emission at photon energies $>$30 
MeV was also observed from Orion, with the EGRET instrument on {\it 
CGRO} (Digel, Hunter, \& Mukherjee 1995). This gamma-ray emission is 
consistent with pion production and bremsstrahlung due to 
irradiation by standard Galactic cosmic rays (Digel et al. 1995). As 
such cosmic rays underproduce the observed line emission by at least 
three orders of magnitude, the gamma-ray line production in Orion 
must predominantly be a low energy cosmic ray phenomenon. 
Information on the spatial distribution of the gamma-ray line 
emission in Orion has come from both the COMPTEL and {\it CGRO}/OSSE 
observations. The extended nature of the emission seen in the 
COMPTEL map of Orion (Bloemen et al. 1997) could provide an 
explanation for the fact that so far it was not possible to confirm 
the COMPTEL results with OSSE (Murphy et al. 1996; Harris et al. 
1998). 

Based on the observed line widths, Bloemen et al. (1994) first 
suggested that the line emission is produced by accelerated C and O 
ions interacting with ambient H and He, rather than by accelerated 
protons and $\alpha$-particles interacting with ambient C and O. 
More detailed analyses of the initial COMPTEL data have shown that a 
mix of the two processes could not be ruled out (Ramaty, Kozlovsky, 
\& Lingenfelter 1995; Cowsik \& Friedlander 1995). But, as the 
emission peaks in the more recent COMPTEL data do not appear at the 
line center energies for $^{12}$C and $^{16}$O de-excitations 
(Bloemen et al. 1997), a significant narrow-line contribution from 
accelerated proton and $\alpha$-particle interactions seems to be 
excluded (Kozlovsky, Ramaty, \& Lingenfelter 1997). This conclusion 
is also supported by energetic arguments, as the very large power 
deposited by the accelerated particles into the ambient medium in 
Orion is lowered by enhancing the C-to-proton and O-to proton 
abundance ratios (Ramaty et al. 1995; Ramaty, Kozlovsky \& 
Lingenfelter 1996). Apart from the observed emission in the 3-7 MeV 
band, the COMPTEL observations revealed only upper limits at other 
gamma-ray energies (Bloemen et al. 1994, 1997). In particular, the 
upper limit on the 1-3 MeV emission sets constraints on the 
accelerated Ne-Fe abundances relative to those of C and O. The 
suppression of both the Ne-Fe and proton and $\alpha$-particle 
abundances relative to C and O could be understood if the seed 
particles injected into an as-yet unknown particle accelerator (see 
Nath \& Biermann 1994; Bykov \& Bloemen 1994) come from the winds of 
massive stars or the ejecta of supernovae resulting from massive 
progenitors (Bykov \& Bloemen 1994; Ramaty et al. 1995; Cass\'e, 
Lehoucq, \& Vangioni-Flam 1995; Ramaty et al. 1996; Parizot, 
Cass\'e, \& Vangioni-Flam 1997a). Ip (1995) and Ramaty et al. (1996) 
have also considered the possible acceleration of ions resulting 
from the breakup of interstellar dust. 

The gamma-ray line production in Orion should be accompanied by a 
large ionization rate of the ambient medium which could exceed the 
observed infrared luminosity (Cowsik \& Friedlander 1995). This 
problem is alleviated if the gamma-rays are produced at cloud 
boundaries, but not in their interiors. The accelerated particles 
could have ionized $\sim$2$\times$10$^4$M$_\odot$ in 10$^5$ years 
(Ramaty 1996), a small fraction of the total available mass. It is 
thus possible that a large fraction of the power that accompanies 
the gamma-ray production is deposited in an ionized gas.

While the X-ray emission produced by low energy particle 
interactions is potentially a promising tracer of low energy 
cosmic rays in the Galaxy (e.g. Hayakawa \& Matsuoka 1964), 
there are as-yet no astrophysical X-ray observations that
unambiguously indicate the presence of such cosmic rays. The 
Orion region, however, has become an interesting target owing to the 
COMPTEL discovery of the nuclear gamma-ray line emission. A variety of 
processes lead to X-ray production by low energy ion 
interactions. Inverse bremsstrahlung (Boldt \& Serlemitsos 
1969) results from the interactions of fast ions and ambient 
electrons; secondary electron bremsstrahlung is produced by 
knock-on electrons accelerated in fast ion interactions 
(Hayakawa \& Matsuoka 1964). Both of these processes lead to 
continuum X-ray emission. X-ray line emission results from 
atomic de-excitations in the fast ions following electron 
capture (Silk \& Steigman 1969; Watson 1976; Pravdo \& Boldt 
1975; Bussard, Ramaty, \& Omidvar 1978) and in ambient ions 
following inner-shell vacancy creation. The latter process has 
not yet been applied to astrophysics. Dogiel et al. (1997) have 
recently considered the X-ray emission that should accompany 
the gamma-ray line production in Orion. They have only 
considered the secondary electron bremsstrahlung and concluded 
that the 0.5-2 keV emission that accompanies the observed 
gamma-ray line emission from Orion will exceed the upper limits 
that they derived using ROSAT observations. We have 
subsequently taken into account both continuum processes and 
line emission from de-excitations in fast O (Ramaty, Kozlovsky, 
\& Tatischeff 1997a) and showed that, even though the inverse 
bremsstrahlung is more important than the secondary electron 
bremsstrahlung, the total X-ray continuum emission from Orion 
is not inconsistent with the Dogiel et al. (1997) derived ROSAT 
upper limit. On the other hand, we showed that a conflict may 
exist between that ROSAT upper limit and the X-ray line 
emission following electron capture onto fast O nuclei. 
However, as we suggested, this conflict could be resolved if 
the X-ray and gamma-ray lines are produced in an ionized medium 
or if the current epoch accelerated particle spectrum is 
suppressed at low energies, for example by energy losses.

In this paper we present detailed calculations of X-ray continuum 
and line production by accelerated particle interactions. The bulk 
of our treatment is for a steady state, thick target model with a 
neutral ambient medium. This is the standard model in which most of 
the gamma-ray calculations have been carried out (e.g. Ramaty et al. 
1996). But we have also investigated the effects of an ionized 
ambient medium and a time-dependent model, as these modifications 
could have important consequences on the predicted X-ray to 
gamma-ray production ratio. In our treatment of the continuum, we 
have supplied the details of the calculations and we have improved 
the employed cross sections, thereby confirming our previous 
preliminary results (Ramaty et al. 1997a). We have greatly expanded 
our treatment of X-ray line emission. We have investigated in detail 
the atomic physics relevant to line emission from de-excitations in 
fast O, checking our theoretical calculations against laboratory 
data whenever available. We then expanded the treatment to the other 
abundant accelerated ions (C, N, Ne, Mg, Si, S and Fe), and we have 
also calculated the X-ray line emission produced in ambient ions 
following inner-shell vacancy creation by the accelerated particles. 
We have used the ROSAT all-sky survey (Snowden et al. 1995) to 
derive the X-ray count rates from the Orion region that could be 
associated with accelerated particle interactions; the implied 
fluxes are quite different from the upper limit given by Dogiel et 
al. (1997). The unambiguous future detection of the predicted X-rays 
produced by accelerated particles in Orion, and potentially 
elsewhere in the Galaxy, should provide important new insights into 
the origin of the low energy cosmic rays whose presence in Orion is 
revealed by the COMPTEL gamma-ray line observations.

\section{Interaction model}

We consider a steady state, thick target interaction model in which 
accelerated particles with given energy spectra and composition are 
injected at a constant rate into an interaction region of solar 
composition and produce atomic and nuclear reactions as they slow 
down to energies below the thresholds of the various reactions. We 
perform the calculations in a neutral ambient medium. In \S 5.1 we 
discuss the expected modifications for a partially ionized medium 
and in \S 5.2 the modifications that could result if the accelerated 
particle population is allowed to evolve in time.

The thick target X-ray and gamma-ray production rates can be written 
as (e.g. Ramaty et al. 1996)
\begin{eqnarray}
Q = {1 \over \big(1 + 1.8 {n_{He} \over n_{H}}\big) m_p} \sum_{ij} 
{A_i} {n_j \over  n_H} \int_0^{\infty} 
{\sigma_{ij}(E_n) dE_n \over Z_i^2(eff) \big({dE \over dx}\big)_{_{p,H}}} 
\int_{E_n}^\infty {dN_i \over dt} (E_n') dE_n'~.
\end{eqnarray}
Here $i$ and $j$ range over the accelerated and ambient particle 
types that contribute to the atomic or nuclear product considered; 
$E_n$ is the accelerated particle energy per nucleon; ${dN_i \over 
dt}(E_n)$ is the differential injection rate of projectiles of type 
$i$ into the target region, measured in particles 
(MeV/nucleon)$^{-1}$s$^{-1}$; $\sigma_{ij}(E_n)$ is the cross 
section for the particular reaction considered; $\big({dE \over dx} 
\big)_{_{p,H}}$ is the proton energy loss rate per g cm$^{-2}$ in 
ambient neutral H; $Z_i$, $A_i$ and $c\beta_i$ are the nuclear 
charge, mass and velocity of projectiles of type $i$; $Z_i(eff) = 
Z_i \big[1 - {\rm exp}(-{137 \beta_i/Z_i^{2/3}})\big]$ is the 
equilibrium ionic charge in a neutral ambient medium (Pierce \& 
Blann 1968); $n_H$, $n_{He}$ and $n_j$ are the densities of ambient 
H, He and constituent $j$; and $m_p$ is the proton mass.

We perform calculations with an accelerated particle source 
spectrum of the form (Ramaty et al. 1996)
\begin{eqnarray} 
{dN_i \over dt} (E_n) \propto E_n^{-1.5}e^{-E_n/E_0}~. 
\end{eqnarray}
The power law $E_n^{-1.5}$ is appropriate for strong shock 
acceleration in the nonrelativistic region. The exponential cutoff, 
introduced by Ellison \& Ramaty (1985) for solar flare acceleration, 
could be caused by a finite shock size or finite acceleration time. 
In the case of Orion, arguments of energetics require a hard 
spectrum in the nonrelativistic region (Ramaty et al. 1996), such as 
that given by equation (2) with as high a value for the turnover 
energy ($E_0$$\lsim$100MeV/nucleon) as allowed by the requirement 
that the high energy gamma-ray emission due to pion decay not be 
overproduced (Tatischeff, Ramaty, \& Mandzhavidze 1997). While such 
values of $E_0$ are reasonable for solar flares, it is not clear 
that in the case of strong shock acceleration the relevant geometry 
and time scales for Orion will lead to a cutoff at a sufficiently 
low energy. Alternatively, the acceleration in the nonrelativistic 
region could be stochastic due to turbulence or an ensemble of weak 
shocks (Bykov \& Fleishman 1992; Bykov 1995). This mechanism 
predicts an even harder nonrelativistic spectrum ($E_n^{-1}$), and 
one which steepens at higher energies where the acceleration is due 
to single weak shocks. Because of its simplicity, we shall use 
equation (2) in our subsequent calculations, allowing $E_0$  to vary 
from 10 to 100 MeV/nucleon.

We employ three different compositions for the accelerated 
particles: CRS and WC (Ramaty et al. 1996, table 1); and OB (Parizot 
et al. 1997a, table 1 the OB/0.02 column). CRS is the composition of 
the current epoch Galactic cosmic-ray sources, WC is the composition 
of the winds of Wolf-Rayet stars of spectral type WC, and OB is the 
average composition of the winds from OB associations. The C and O 
abundances relative to protons, $\alpha$-particles and heavier 
metals are much higher for the WC case than those for the CRS case. 
For the OB case, these abundances ratios are intermediate between 
those of the CRS and WC cases. 

\section{X-ray continuum emission}

Energetic ions produce continuum X-ray emission via both inverse 
bremsstrahlung (Boldt and Serlemitsos 1969, hereafter IB) and 
bremsstrahlung from secondary knock-on electrons (see also Anholt 
1985). For nonrelativistic protons interacting in stationary H, the 
inverse bremsstrahlung cross section is identical to the 
bremsstrahlung cross section for electrons of kinetic energy 
$(m_e/m_p)E$ also interacting in stationary H; here $m_e$ and $m_p$ 
are the electron and proton masses respectively and $E$ is the 
proton kinetic energy. We used equation 3BN from Koch and Motz 
(1959). For heavier ions interacting in an ambient medium consisting 
of heavier atoms, we replaced the proton energy $E$ by the energy 
per nucleon of the projectile, $E_n$, and multiplied the cross 
section by $Z_i^2$$Z_j$, where $Z_i$ and $Z_j$ are the fast ion and 
ambient atom charge numbers, respectively.

The secondary electrons are knock-on electrons which subsequently 
produce bremsstrahlung by interacting with the ambient atoms. The 
angle averaged differential X-ray production cross section for 
secondary electron bremsstrahlung (SEB) can be written as
\begin{eqnarray} 
{d\sigma \over d\epsilon}(E_n,\epsilon)={Z_i^2 
Z_j^2 \over m_p} \int_{\epsilon}^{\infty} {d\sigma_{BR} \over 
d\epsilon} (E_e,\epsilon) {dE_e \over ({A_j \over Z_j}) {dE_e \over 
dx}(E_e)    } \int_{E_e}^{\infty} {d\sigma_{KN} \over 
dE_e}(E_n,E_e')dE_e'~, 
\end{eqnarray}
where $(d\sigma_{KN}/dE_e)(E_n,E_e')$ is the knock-on cross section 
for the ejection of an electron of energy $E_e'$ by a proton of 
energy $E_n$ interacting with H (Chu et al. 1981 with corrections at 
high energies from Rudd et al. 1966 and Toburen and Wilson 1972); 
$(d\sigma_{BR}/d\epsilon)(E_e,\epsilon)$ is equation 3BN of Koch and 
Motz (1959) for electrons of energy $E_e$ interacting with H and 
radiating a photon of energy $\epsilon$; and $dE_e/dx$ is the 
electron energy loss rate per g cm$^{-2}$ in the ambient medium 
(Berger and Seltzer 1982). Since $dE_e/dx$ scales approximately as 
$Z_j$/$A_j$, where $A_j$ is the target atomic number, the overall 
cross section scales as $Z_i^2$$Z_j^2$. In comparison, the IB cross 
section scales as $Z_i^2$$Z_j$. 

In Figure~1a we show our calculated IB and SEB cross sections for 
protons interacting in H and Be, and compare our results for Be with 
experimental data (Chu et al. 1981). We see that while for the Be 
target the two calculated cross sections are approximately equal, 
for the H target the IB cross section dominates (except at the 
highest energies). This is the direct consequence of the 
dependencies of the cross sections on the target charge number $Z_j$ 
derived above. In the case of the Be target we get good 
agreement with the experimental data at 90$^\circ$, the discrepancy 
of less than a factor of 2 possibly being due to our angle averaging 
and contaminations in the experiment, for example Compton scattering 
of gamma rays in the Be target. 

Using equation(1) we calculated the ratio of the IB and SEB 
productions as a function of photon energy for an ambient medium 
with solar abundances, for various values of the characteristic 
energy $E_0$ [Eq.~(2)], and various accelerated particle 
compositions. We find that, as expected, the ratio is practically 
independent of the accelerated particle composition. We thus show 
results only for the CRS composition (Figure~1b). We see that, 
except at the highest energies, the IB-to-SEB ratio ranges from 
about unity to almost 10, depending on the photon energy and the 
spectrum of the accelerated particles (i.e. the value of $E_0$).

\section{X-ray line emission}

X-ray line emission results mainly from 2$p$ to 1$s$ and 3$p$ to 1$s$ 
transitions, in either the fast ions or the ambient atoms, giving 
rise to K$\alpha$ and K$\beta$ X-rays respectively. In the case of 
the fast ions, the 2$p$ and 3$p$ states can be populated either by 
electron capture from ambient atoms (i.e. charge exchange) or by 
excitation of 1$s$ electrons for ions having one or two electrons. We 
neglected  the K$\alpha$ and K$\beta$ X-ray production from fast 
ions having more than two electrons as the collision energy in this 
case is mostly given to the outer electrons and thus not giving rise 
to a significant K-shell electron excitation. The fast ions also 
produce K-shell vacancies in the ambient atoms (Garcia, Fortner, \& 
Kavanagh 1973; Cahill 1980). We first consider the K X-ray production 
in fast O.

\subsection{K X-rays from fast oxygen}

X-ray line production from fast O has already been treated by 
Pravdo \& Boldt (1975) and Watson (1976). The latter has considered 
the interaction of energetic O with only ambient hydrogen. But, 
as shown by Bussard et al. (1978) for K$\alpha$ X-ray emission from 
fast Fe, heavier ambient elements cannot be neglected despite their 
lower abundances because of the strong dependence of the 
charge-exchange cross section on the charge of the target nucleus. 
Pravdo \& Boldt (1975) have taken into account the contributions of 
the heavier elements, but as we shall see, their use of the 
Oppenheimer-Brinkman-Kramers (OBK) approximation for the 
charge-exchange cross section in the form presented by Schiff (1954) 
gives a poor description of the available experimental data. 

The main lines produced by fast O ions result from the 2$p$ to 
1$s$ transitions in H- and He-like projectiles giving rise to 
K$\alpha$ X-rays at 0.65 keV (O{\sml \ VIII} line) and 0.57 keV 
(O{\sml \ VII} line), respectively (Matthews et al. 1973). In 
addition, we considered the 3$p$ to 1$s$ transitions, which 
give rise to non-negligible K$\beta$ X-rays at 0.77 and 0.67 
keV, for H- and He-like O ions, respectively (Hopkins et al. 
1974). 

\subsubsection{Charge Exchange cross section}

Even though the OBK cross section is known to exceed the 
experimental data by a factor of 10 or more in certain cases, it 
predicts quite accurately the shape of the measured cross sections 
versus energy (e.g. the review of Belki\'c, Gayet, \& Salin 1979). 
Schiff (1954) has derived a general form for the OBK cross section 
for capture of electrons by fast ions from the ground state of a 
H-like ion. This formula is not appropriate to describe 
electron capture from heavy targets containing many atomic 
electrons. Nikolaev (1967) has extended the OBK cross section to 
heavy ion charge exchange, taking into account the complete atomic 
configuration of the target atoms. He also derived a semi empirical 
scaling relation in order to adapt his theoretical result to 
experimental data. But subsequent measurements have shown that a 
discrepancy of a factor of 2-3 remains between theory and experiment 
in some cases (e.g. Ferguson et al. 1973). Thus, instead of using 
the Nikolaev (1967) semi empirical factor, we normalized the 
Nikolaev cross section to experimental data. In the Nikolaev 
formalism we used the binding energies of the target electrons from 
Sevier (1979), and the Slater (1930) rules to take into account the 
external screening of the nuclear charge by the inner shells. 

A compilation of experimental cross sections is shown in Figure~2, 
together with theoretical calculations. The closed symbols show K 
X-ray production cross sections due to charge exchange for fast 
fully stripped O and F interacting with neutral H, He, Ar and Kr. 
The solid curves are our fits to these data obtained by using the 
Nikolaev (1967) formalism to calculate the capture cross sections 
$\sigma_{n\geq2}$ to all states with principal quantum number 
$n\geq2$. To take into account the captures which populate the 2$s$ 
metastable state (which does not lead to K X rays), we reduced 
$\sigma_{n\geq2}$ by 4\% based on the calculations of Guffey, 
Ellsworth \& Macdonald (1977) for O and F projectiles in He. This 
small correction, however, is negligible in comparison with the 
overall normalizing factor of 0.1 that we applied to the calculated 
curves to fit the data. The open symbols show the total electron 
capture cross section from the charge spectrometer experiment of 
Macdonald et al. (1972) for fluorine-nitrogen collisions. The dashed 
curve is our fit to these data, again using the Nikolaev (1967) 
formalism but with a sum over all n$\geq$1, and the same normalizing 
factor of 0.1. It is encouraging that with a single normalizing 
factor we could account for all of the data in Figure~3 with 
accuracy better than a factor of 2. In addition, the normalizing 
factor of 0.1 is within the range of the Nikolaev semiempirical 
normalizations for charge exchange on fast protons. 

We also compared the predictions of the Schiff (1954) formula 
with the data shown in Figure~2. While for the H target, the 
formula gives an acceptable fit with the same normalization of 
0.1, it predicts incorrect spectral dependencies for targets 
heavier than He, which is not surprising since the formula was 
developed for H-like targets in the ground state.

\subsubsection{Excitation cross section}

The plane-wave-Born-approximation (PWBA) gives a satisfactory 
description for the Coulomb excitation of atomic electrons 
(Bates 1962). Figure~3 shows a comparison of theoretical cross 
sections with the experimental data of Hopkins, Little, \& Cue 
(1976a) for the excitation of 1$s$ electrons which leads to K 
X-ray emission. The theoretical curves consist of the sum of 
the 1$s$$\rightarrow$2$p$ and 1$s$$\rightarrow$3$p$ excitation 
cross sections, excitations to higher levels, which have much 
smaller cross sections, being neglected. The cross sections for 
the excitation of F$^{7+}$ are greater than twice those of 
F$^{8+}$ (F$^{7+}$ has two K-shell electrons which could be 
excited), because we took into account the screening of the 
nuclear charge of the projectile by the other K-shell electron. 
Although the PWBA theory indicates that the excitation cross 
sections scale as the target charge squared, the Hopkins et al. 
(1976a) measurements show that the cross sections for 
excitation by He are $\sim$20\% lower than four times the 
corresponding ones for excitation by H. This discrepancy was 
attributed to the screening effect of the helium electrons 
(Hopkins et al. 1976a). To take this effect into account, we 
used a normalization factor of 0.8 for all the target elements 
heavier than hydrogen in addition to the scaling with the 
target charge squared. This approximation has only a small 
impact on the final result, since the contribution of the 
elements heavier than He to the excitation represents less than 
4\% of the final X-ray production.  We neglected the 
contributions of excitation reactions in which both the 
projectile and the target are excited based on the results of 
Moiseiwitsch \& Stewart (1954). 

\subsubsection{Equilibrium charge fractions}

The O ions lose energy more slowly than they capture and lose 
electrons by interacting with a neutral ambient medium of solar 
composition  (see the discussion of Bussard et al. 1978 for fast Fe 
ions). Thus, in a steady state, the charge fractions $F_{n}$, where 
$n$ is the number of bound electrons of the projectile, satisfy the 
equation 
\begin{eqnarray} 
F_{n}\sigma_{c}^{(n)}=F_{n+1}\sigma_{I}^{(n+1)}~, 
\end{eqnarray} 
where $\sigma_{c}^{(n)}$ and $\sigma_{I}^{(n)}$ are, respectively, the 
electron capture and ionization cross sections for O ions 
with $n$ electrons interacting with an ambient medium of solar 
composition. Ionization cross sections scale approximately as the 
target charge squared (Rule 1977). Thus, because of the low 
abundances of the heavier elements (Anders \& Grevesse 1989), the 
fast ions lose their electrons by colliding mainly with ambient H 
and He. This is not the case for electron captures because of the 
strong dependence of the charge exchange cross section on the charge 
of the target nucleus (e.g. Figure~2). We took into account the 
ambient H, He, C, N, O, Ne, Mg, Si, S and Fe to calculate the total 
electron capture of the fast O ions in the medium of solar 
composition. Ionization cross sections of O ions in collisions 
with neutral H and He were taken respectively from Watson (1976) and 
Rule (1977). The electron capture and loss cross sections for fast O 
in a medium of solar composition are shown in Figure~4a and the 
$F_{0}$, $F_{1}$ and $F_{2}$ charge fractions determined from 
equation (4) are shown in Figure~4b. We see that contrary to the 
charge exchange cross sections, the ionization cross sections have a 
weak energy dependence. As most of the X-ray line emission is 
produced near the energy at which the charge exchange and ionization 
cross sections are equal, and because of the weak energy dependence 
of the ionization cross sections, the final X-ray line production is 
not very sensitive to the exact magnitude of the charge exchange 
cross section. We found that an uncertainty of a factor of 2 in the 
charge exchange cross section (\S 4.1.1) leads to only a 25\% 
uncertainty in the X-ray emissivity.

\subsubsection{K X-ray multiplicities}

The multiplicities of the H-like (O{\sml \ VIII}) and He-like 
(O{\sml \ VII}) K$\alpha$ lines produced by a fast O of initial 
energy $E_{n0}$ slowing down in a medium with solar abundances due 
to interactions with the ambient constituent $j$ of abundance 
$n_j$/$n_H$ can be obtained from equation (1), 
\begin{eqnarray} 
M_{\rm O{\sml \ VIII}}(j)={A_i \over \big(1 + 1.8 {n_{He} \over n_{H}}\big) m_p} 
\Big({n_j \over  n_H}\Big) \int_0^{E_{n0}} {dE_n \over Z_i^2(eff) 
\big({dE \over dx}\big)_{_{p,H}}} \Big(F_0 \sigma_{c;j}^{(0)}(2p) + 
F_1 \sigma_{ex;j}^{(1)}(1s\rightarrow2p)\Big) 
\end{eqnarray} 
\noindent and
\begin{eqnarray} 
M_{\rm O{\sml \ VII}}(j)={A_i \over \big(1 + 1.8 {n_{He} \over n_{H}}\big) m_p} 
\Big({n_j \over  n_H}\Big) \int_0^{E_{n0}} {dE_n \over Z_i^2(eff) 
\big({dE \over dx}\big)_{_{p,H}}} \Big(F_1 \sigma_{c;j}^{(1)}(2p) + 
F_2 \sigma_{ex;j}^{(2)}(1s\rightarrow2p)\Big)~. 
\end{eqnarray} 
Here the $F_n$ are the charge fractions of the O ions; 
$\sigma_{c;j}^{(n)}(2p)$ is the electron capture cross section into 
the 2$p$ level, either directly or following a cascade, by an ion 
which already has $n$ electrons from the target element $j$; 
$\sigma_{ex;j}^{(n)}(1s\rightarrow2p)$ is the 1$s$$\rightarrow$2$p$ 
excitation cross section for an O ion having $n$ electrons by 
the target element $j$. The O K$\beta$ X-ray multiplicities are 
obtained by replacing 2$p$ by 3$p$ in equations (5) and (6). 

To calculate $\sigma_{c;j}^{(n)}(2p)$ and $\sigma_{c;j}^{(n)}(3p)$, 
we first calculated the total K X-ray production as in \S~4.1.1, and 
then used the experimental result of Hopkins et al. (1974) to 
determine the relative contributions of the K$\alpha$ and the 
K$\beta$ lines. Hopkins et al. (1974) measured the K X-ray 
production from fully stripped O of 1.9 MeV/nucleon interacting with 
various targets. As the projectiles were initially devoid of 
electrons, the detected K X-ray emission could have arisen only from 
electron capture. They found that, independent of the target charge, 
about 60\% of the total K X-ray production was in the K$\alpha$ line 
and 20\% in the K$\beta$ line, the remaining 20\% coming from the 
rest of the series. Combining this result with the calculation of 
$\sigma_{ex;j}^{(n)}(1s\rightarrow2p)$ and 
$\sigma_{ex;j}^{(n)}(1s\rightarrow3p)$, we eventually found that 
each of the two K$\beta$ lines, i.e. from the H- and 
He-like O, represents 1/4 of the corresponding K$\alpha$ 
line. 

We have evaluated equations (5) and (6) and the results are given in 
Tables 1 and 2, where we have separated the contributions from 
electron captures and excitations of 1$s$ electrons. The 
multiplicities are independent of the initial O energy for 
$E_{n0}$$\gsim$10 MeV/nucleon, because at these high values of 
$E_{n0}$ the nuclei are fully stripped and the probability for 
capturing electrons is very small. The results of Tables 1 and 2 can 
be compared with the multiplicities given by Pravdo \& Boldt (1975). 
Although our results for the contribution of charge exchange are 
$\sim$4 times lower than theirs, the total K$\alpha$ X-ray 
production is only 1.8 times lower. This is because excitation is 
now seen to be the dominant mechanism. This result is supported by 
the accelerator measurement of Hopkins et al. (1976a), which provides 
a clear signature that excitation is the dominant process of K X-ray 
production for 1-3 MeV/nucleon F$^{7+}$ and F$^{8+}$ projectiles in  
hydrogen and helium. The same conclusion was found also by Bussard 
et al. (1978) for the K$\alpha$ X-ray emission from fast Fe. In 
particular, we can see from Table~2 that the excitation contribution 
to the O{\sml \ VII} line production, which was not taken into 
account by Pravdo \& Boldt (1975), represents 60\% of the total 
multiplicity. 

Using Equations (5) and (6) we can also evaluate the 
differential multiplicities $\sum_{j} dM_{\rm O{\sml \ 
VIII}}(j)/dE_{n0}$ and $\sum_{j} dM_{\rm O{\sml \ 
VII}}(j)/dE_{n0}$, i.e. the number of photons produced by the 
projectile, due to interactions with all the constituents of 
the ambient medium, as it slows down over the differential 
energy interval $dE_n$. The results for O are shown in Table~3 
and Figure~5c. Similar multiplicities for C, N, Ne, Mg, Si, S 
and Fe are also shown in Tables~3 and 4 and in Figures~5 and 6. 
These multiplicities are discussed below. We see that for O the 
differential multiplicities peak around 1 MeV/nucleon and fall 
off rapidly at both lower and higher energies. The differential 
multiplicities for O can be compared with the results of Pravdo 
\& Boldt (1975). The positions of the peaks are consistent 
although, as already discussed, the magnitudes of the 
multiplicities are lower. 

\subsection{K X-rays from the other fast ions}

The major X-ray lines from the fast ions are given in Table 5. We 
used the same formalisms for the charge exchange and the excitation 
cross sections as described for O. In particular, we used for the 
charge exchange cross sections the same normalization factor of 0.1 
for all the projectiles and targets (\S~4.1.1). We checked our 
charge exchange calculations with the experimental data of Berkner 
et al. (1977) for 3.4 MeV/nucleon Fe projectiles in charge states 
+20 to +25 interacting with H$_2$ and found agreement within 40\%. 
For the ionization cross sections, we used the scaling relation 
(Rule 1977)
\begin{eqnarray} 
\sigma_{I}^{Z_i}(v_i)=(8/Z_i)^{4} \sigma_{I}^{8}(8v_i/Z_i)~, 
\end{eqnarray} 
where $v_i$ and $Z_i$ are the speed and charge of the projectile, 
and $\sigma_{I}^{8}$ is the ionization cross section of fast O ions 
in an ambient medium of solar composition (\S 4.1.3, and Figure~4a). 
We assumed for all the fast ions that the K$\alpha$ and the K$\beta$ 
lines represent respectively 60\% and 20\% of the K X-ray production 
from electron capture. 

The results for the K$\alpha$ differential multiplicities, 
along with the total multiplicities, are shown in Figures~5 and 
6. As for O, we found that the multiplicities for the K$\beta$ 
photons are approximately 1/4 of the corresponding K$\alpha$ 
multiplicities. The multiplicities decrease with increasing 
projectile charge, from a total multiplicity of 87 
photons/projectile for C to 7 photons/projectile for Fe. The 
projectile energies at which the  differential multiplicities 
peak increase with increasing projectile charge. Our results 
for Fe are in good agreement with the previous calculations of 
Bussard et al. (1978). In particular, we agree that excitation 
of 1$s$ electrons is the dominant mechanism of K$\alpha$ X-ray 
production, that ambient O and Fe are the main contributors to 
electron capture by fast Fe ions and that the peak of the X-ray 
line production from fast Fe is at $\sim$9 MeV/nucleon. Both 
calculations of the total emissivity of Fe K$\alpha$ lines 
agree within 20\% (see the table 2 and the notes added in the 
manuscript of Bussard et al. 1978). 

\subsection{K X-rays from the ambient atoms}

The K X-ray line emission from the ambient atoms results from 
the filling of inner-shell vacancies produced by the fast ions. 
In the case of proton impact, the line energies correspond to 
the difference of the orbital energies, because both the 
vacancy production and the subsequent filling of the vacancy 
occur in times short compared to the relaxation times for the 
atomic wave functions (Garcia et al. 1973). For heavy-ion 
collisions, the lines could be shifted by several tens of eV, 
significantly broadened and slitted up into several components, 
due to multiple inner-shell plus outer-shell simultaneous 
ionizations (Garcia et al. 1973). We have not taken into 
account these effects, although we believe that they could 
provide interesting constraints on the accelerated particle 
composition through fine spectroscopy. We considered the 
K$\alpha$ and K$\beta$ lines from ambient C, N, O, Ne, Mg, Si, 
S, Ar, S and Fe and have determined the transition energies 
(Table~6) from the atomic electron binding energy tables of 
Sevier (1979). We do not consider K$\beta$ X-rays from C, N, O, 
Ne and Mg, as these atoms in their ground state do not have 
3$p$ electrons. 

The X-ray line production cross section can be written as
\begin{eqnarray} 
\sigma_{X}=\sigma_{I}wk, 
\end{eqnarray} 
where $\sigma_{I}$ is the cross section for the collisional 
ionization leading to the K-shell vacancy, $w$ is the 
fluorescence yield for the K shell (Krause 1979) and $k$ is the 
relative line intensity among the possible transitions which 
can fill the inner-shell vacancy (Salem, Panossian, \& Krause 
1974). For proton impact, we calculated the ionization cross 
sections from the semiempirical formula of Johansson \& 
Johansson (1976), with an extrapolation at high energy from 
appendix 2 of Garcia et al. (1973). For heavier 
projectiles, we assumed a Z$_{i}^2$ dependence, as predicted 
by both the PWBA and the impulse approximation theories. This 
assumption gives a poor description of the available experimental 
data at low energies due to molecular effects, but becomes 
reasonable as the cross section reaches its maximum, i.e. for 
$E_n \simeq (m_p/m_e) u_K$ (in MeV/nucleon), where $u_K$ is the 
binding energy of the K shell  (Garcia et al. 1973). 

K X-ray production cross sections for fast protons colliding with C, 
O, Si and Fe are shown in Figure~7. The K$\alpha$ line production 
cross sections do not show a strong dependence on the target charge 
even though the ionization cross section varies as $Z_j^{-4}$ 
(Johansson \& Johansson 1976), because this variation is partially 
compensated by the $Z_j^{3.25}$ dependence of the fluorescence 
yields up to Fe (Krause 1979). 

\section{Results and Applications for Orion}

We developed a code for the calculation of the X-ray emission 
resulting from the various continuum and line producing processes 
discussed above. The code allows the use of various accelerated 
particle energy spectra and compositions in a steady state, thick 
target model. The ambient medium is assumed to be neutral with a 
solar composition. We discuss in \S~5.1 the modifications that would 
result for X-ray production in a partially ionized medium and in \S 
5.2 those for a time-dependent model. We took into account the 
photoelectric absorption of the X-rays, again using solar 
abundances, and the cross sections of Morrison \& McCammon (1983). 
We present results normalized to the accompanying 3-7 MeV nuclear 
gamma-ray line emission which we calculated as in Ramaty et al. 
(1996).

The calculated X-ray fluxes, normalized to a 3-7 MeV nuclear 
gamma-ray flux of 10$^{-4}$ photons cm$^{-2}$ s$^{-1}$ (the 
approximate value of the observed gamma-ray line emission from 
Orion), are shown in Figures 8 and 9 for the CRS and WC 
compositions, respectively. We took into account photoelectric 
absorption using N$_H$$=$3$\times$10$^{21}$ cm$^{-2}$, our 
estimate for the H column density towards Orion. We estimated 
this column density from the H{\sml I} survey of Heiles \& 
Habing (1974) and the CO map of Dame et al. (1987), together 
with a  CO-to-H$_{2}$ conversion factor of 
(1.06$\pm$0.14)$\times$10$^{20}$ cm$^{-2}$ (K km 
s$^{-1}$)$^{-1}$ (Digel et al. 1995).

We first note that the continuum emission is indeed dominated 
by the inverse bremsstrahlung (\S~3). We have calculated the 
SEB differential luminosity for $E_0$=30 MeV/nucleon without 
including the photoelectric absorption and compared the results 
with those given in figure 1 of Dogiel et al. (1997). For the 
CRS composition our results are lower than theirs by factors of 
19, 6.6 and 1.9 at $E_{\rm x}$=0.2, 1 and 10 keV, respectively; 
for the WC composition, the discrepancy is even larger, the 
corresponding factors being 75, 31 and 9. The discrepancy is 
due, in part, to the approximate normalization to the nuclear 
gamma-ray line emission used by Dogiel et al. (1997). However, 
by repeating their  calculation using the same formalism and 
input data (Hayakawa 1969), we find results which agree with 
our more accurate approach but which are lower than that of 
Dogiel et al. (1997) by about an order of magnitude at 1 keV.

We calculated the width of the lines produced by the fast ions 
assuming an isotropic angular distribution for these particles. We 
see that below about 1 keV the X-ray emission is dominated by these 
lines, in particular the lines from fast C and O. For the much 
narrower lines from the ambient atoms, we assumed delta functions at 
the nominal line energies (Table~6) and plotted the spectra in 10 eV 
bins. The most prominent narrow line is that at 6.4 keV from ambient 
Fe. The width of the lines resulting from de-excitations in the fast 
ions are quite broad. For example, the O lines, which are produced 
around 1 MeV/nucleon, have widths of about 0.06 keV (FWHM). This 
large width distinguishes them from the X-ray lines produced in a 
hot plasma.

By comparing Figures~8 and 9, we see that the line-to-continuum 
ratio below 1 keV is lower for the CRS composition than for the WC 
composition because of the additional bremsstrahlung from the 
protons and $\alpha$-particles. The line-to-continuum ratio below 1 
keV is also lower for the higher $E_0$, because the fast ions 
produce line emission only at the low energies (Figures~5 and 6). 
The X-ray emission at high energies ($\gsim$10 keV) is very 
sensitive to the value of $E_0$. 

In Figures 10 and 11 we compare the expected X-ray fluxes from Orion 
with the diffuse extragalactic X-ray background. We calculate the 
X-ray emission for the CRS composition with $E_0$=100 MeV/nucleon 
and for the WC composition with $E_0$=20 MeV/nucleon. We consider 
two cases: uniform gamma-ray emission from the box 
[201$^{\circ}$$\leq$l$^{II}$$\leq$217.5$^{\circ}$, 
-21$^{\circ}$$\leq$b$^{II}$$\leq$-9$^{\circ}$] which contains almost 
all of the 3-7 MeV emission observed with COMPTEL, 
(12.8$\pm$1.5)$\times$10$^{-5}$ photons cm$^{-2}$ s$^{-1}$ (Bloemen 
et al. 1997); the hot spot observed with COMPTEL at 
l$^{II}$=213.6$^{\circ}$, b$^{II}$=-15.7$^{\circ}$ from which the 
3-7 MeV emission flux is (3.3$\pm$0.8)$\times$10$^{-5}$ photons 
cm$^{-2}$ s$^{-1}$ (H. Bloemen, private communication, 1996). As the 
size of the spot is not well known, we considered a 9 degree$^2$ box 
[212.25$^{\circ}$$\leq$l$^{II}$$\leq$215.25$^{\circ}$, 
-17.25$^{\circ}$$\leq$b$^{II}$$\leq$-14.25$^{\circ}$] which 
represents the approximate extent of the hot spot in the published 
COMPTEL map (Bloemen et al. 1997). We took the extragalactic 
diffuse X-ray background from Gendreau et al. (1995) below 10 keV 
and from Gruber (1992) at higher energies. We increased the Gruber 
(1992) data by 9\% to achieve a smooth transition to the Gendreau et 
al. (1995) data. 

Considering first the X-ray continuum at high energies, we see that 
if the gamma-ray source is uniformly spread over the entire 200 
degree$^2$ box (Figures 10a and 11a), it will be very difficult to 
observe these X-rays because the predicted emission is much lower 
than the extragalactic background. In the case of the 9 degree$^2$ 
hot spot, the chances of detecting the high energy X-ray continuum 
above the extragalactic background are better, in particular for the 
CRS composition with $E_0$=100 MeV/nucleon (Figure~10b) which yields 
the highest X-ray continuum for a given nuclear gamma-ray flux. On 
the other hand, the predicted line emission, in particular the lines 
from fast O, exceeds the extragalactic background independent of the 
spectrum and composition of the accelerated particles. 

ROSAT observations can be used to test our predictions of the 
expected line emission below 2 keV. We used the ROSAT all-sky survey 
(Snowden et al. 1995) to compare our calculations with the observed 
count rate in the ROSAT R45 energy band, i.e. from 0.47 to 1.2 keV 
(Snowden et al. 1994). We considered the same two source regions as 
in Figures~10 and 11: the 200 degree$^2$ box for which we assume a 
uniform gamma-ray emission; and the 9 degree$^2$ gamma-ray hot spot. 
For the 200 degree$^2$ box ROSAT/PSPC detected a total of 70.1 
counts s$^{-1}$ in the R45 band. We subtracted the contributions of 
the Orion nebula region and the Ori OB1 association, which were 
interpreted as thermal emission most likely from T Tauri type stars 
(Yamauchi, Koyama, \& Inda-Koide 1994; Yamauchi et al. 1996). The 
remaining count rate of 60.6 counts s$^{-1}$ is plotted in 
Figure~12a. For the 9 degree$^2$ hot spot there are 2.7 counts 
s$^{-1}$ in the ROSAT/PSPC R45 band (Figure~12b). 

In the 200 degree$^2$ box, the corresponding X-ray luminosity is 
$\sim$2$\times$10$^{34}$ ergs s$^{-1}$ assuming a distance to Orion 
of 450 pc. This luminosity is 20 times greater than the upper limit 
derived by Dogiel et al. (1997), even though our value pertains to 
the 0.47-1.2 keV range whereas that of Dogiel et al. (1997) is for 
X-rays between 0.5 and 2 keV. We believe that the discrepancy is due 
to the fact that Dogiel et al. (1997) searched for only nonthermal 
continuum whereas we considered all the emission that could not be 
related to known sources in Orion (see above). In particular, 
because of the presence of the lines (Figures~8 and 9), the total 
X-ray emission from low energy particle interactions may look 
similar to thermal emission (Raymond \& Smith 1977) given the rather 
poor energy resolution of ROSAT/PSPC.  

Also shown in Figure~12 are the predicted R45 count rates obtained 
by convoluting the calculated X-ray spectra with the ROSAT/PSPC 
response function (Snowden et al. 1994) and taking into account 
photoelectric absorption with the same H column density of 
3$\times$10$^{21}$ cm$^{-2}$ for the two source regions. The 
results, normalized to the gamma-ray intensities,  are given as 
functions of the characteristic energy $E_0$ [Eq.~(2)] for the CRS, 
OB and WC compositions. The count rates due to extragalactic 
background were obtained in the same manner, i.e. by convoluting the 
spectrum given by Gendreau et al. (1995) with the ROSAT/PSPC 
response function and using the same H column density for the 
photoelectric absorption as for the two Orion source regions. 

In Figure~12 the increase of the predicted count rates as the 
accelerated particle spectrum becomes softer (i.e. with decreasing 
$E_0$) is due to the decrease of the gamma-ray production relative 
to the X-ray production, which occurs at much lower energies than 
the gamma-ray production. The CRS composition provides the lowest 
X-ray emission because of its highest proton and $\alpha$-particle 
abundances relative to those of $^{12}$C and $^{16}$O. Indeed, 
accelerated protons and $\alpha$-particles produce 3-7 MeV emission 
while interacting with ambient CNO, but no significant X-ray line 
emission in the ROSAT energy range. We note that the extragalactic 
X-ray background accounts for only 1/6 of the observed count rates 
in both Figures~12a and 12b. We see in Figure~12a that except for 
the very low $E_0$, the calculated count rates are lower than the 
observed rate, suggesting a contribution of other sources, for 
example T Tauri stars. On the other hand, we see in Figure~12b that 
the predicted X-ray counterpart of the observed nuclear gamma-ray 
line emission from the 9 degree$^2$ hot spot exceeds the ROSAT count 
rate except at the highest allowed values of $E_0$ near 100 
MeV/nucleon. This discrepancy could be mitigated by assuming that 
the area of the gamma-ray hot spot is larger than 9 degree$^2$, 
although for a smaller area a more severe discrepancy would ensue. 
The discrepancy could be resolved if the X-rays and gamma rays are 
produced by accelerated particles interacting in a partially ionized 
medium, for example at cloud boundaries (Bykov \& Bloemen 1994; 
Ramaty et al. 1997a,b; Kozlovsky et al. 1997; Parizot et al. 1997a), 
or in a time-dependent model in which the spectrum of the 
accelerated particles has become quite flat because of energy 
losses. We elaborate on these two effects in the next two 
subsections.

\subsection{Modifications due to a partially ionized ambient medium}

We consider the effects of a partially ionized ambient medium on the 
K X-ray production in the ROSAT R45 energy band (0.47 to 1.2 keV). 
In this energy range, the K X-rays are produced by fast C, N, O and 
Ne. As the nature of the cosmic ray ionization at the cloud 
boundaries is not well understood, we assume for simplicity that the 
ambient H and He are fully ionized but that the heavier ambient 
elements remain essentially neutral. Thus the fast ions capture 
electrons from ambient C and heavier atoms only. As the capture 
cross sections increase rapidly toward lower energies while the 
electron loss cross sections are essentially constant (Figure~4a), 
the charge state distribution of the fast ions in an ionized medium 
is quite similar to that in a neutral medium (Figure~4b), except 
that it is established at lower values of $E_n$. Consequently, K 
X-rays are also produced at lower energies (about 0.1-1 MeV/nucleon 
for the C, N, O and Ne K X-rays) in an ionized medium than in a 
neutral medium. Although the fast ions do not capture electrons from 
ambient H and He, the rate of electron capture is not very different 
from that in a neutral medium because of the more efficient electron 
capture from heavier ambient atoms caused by the increase of the 
capture cross sections toward lower energies (Figure~4a). We also do 
not expect substantial reductions of the magnitude of the K X-ray 
multiplicities due to excitation, because the excitations are mainly 
due to the ambient nuclei and the cross sections are only slowly 
decreasing at lower energies (Figure~3). 

The main difference between K X-ray production from fast ions 
interacting in an ionized or a neutral medium is due to the energy 
losses. It is well-known that fast ions lose more energy in a plasma 
than in a neutral medium, because the collective long-range Coulomb 
interactions are more efficient than ionization. Both the X-ray and 
gamma-ray production rates are reduced, but as we shall see, the 
effect is stronger for the X-rays.

The energy loss rate depends on the effective charge of the fast 
ion. We calculated the effective charge of a fast O in a medium of 
solar composition in which H and He are fully ionized and the result 
is shown by the dashed curve in Figure~13a. We found that this 
effective charge could be approximated by a formula similar to that 
used to describe the effective charge in a neutral medium (\S 2),
\begin{eqnarray} 
Z_i(eff) = Z_i \big[1 - {\rm exp}(-{220 \beta_i/Z_i^{2/3}})\big], 
\end{eqnarray} 
with a modified exponential to take into account the increase of the 
fast ion charge in the ionized medium. We then calculated the O ion 
energy loss in a plasma using equations (4.21) and (4.22) from 
Mannheim and Schlickeiser (1994), in which we substituted the 
nuclear charge of the projectile by its effective charge [Eq. (9)]. 
We assumed a typical temperature of 10$^{4}$ K at cloud boundaries. 
The energy loss is not very sensitive to temperature for 
$E_n\geq0.1$ MeV/nucleon. In Figure~13b we show the ratio of the 
energy loss in an ionized medium to that in a neutral medium for 
fast O ions. We see that while at high energies the increase in the 
energy loss rate is about a factor of 2, below an MeV/nucleon the 
energy loss rate in an ionized medium exceeds that in a neutral 
medium by more than a factor of 6. This result is supported by 
the laboratory measurements of Hoffmann et al. (1994) for the 
energy loss of various fast ions in a hydrogen plasma.

Because the X-ray line emission is produced between about 0.1 and 1 
MeV/nucleon, while the 3-7 MeV gamma-ray line production occurs 
essentially above 10 MeV/nucleon (Ramaty et al. 1996), the ratio of 
thick target [Eq.~(1)] X-ray to gamma-ray line production in an 
ionized medium will be lower than in a neutral medium, by about a 
factor of 3. We emphasize that this decrease is mainly due to the 
increase in the fast ion energy loss rate rather than the changes in 
the electron capture, ionization and excitation.

\subsection{Modifications due to a time-dependent accelerated 
particle population}

Because the K X-ray production in the ROSAT R45 energy band (0.47 to 
1.2 keV) is due to very low energy ions, time-dependent effects 
could significantly reduce the number of such ions relative to that 
of the higher energy gamma-ray line producing ions, thereby 
decreasing the X-ray production rate relative to the gamma-ray 
production rate. 

To illustrate the effect, we first calculate the differential 
equilibrium particle number $Y(E)$ for steady 
state, thick target interactions. This is given by (see Eq.~1)
\begin{eqnarray} 
Y(E) = \Big({ dE \over dt}\Big)^{-1} 
\int_{E}^\infty {dN \over dt} (E') dE'~.
\end{eqnarray} 
where the energy loss rate in a neutral medium is 
\begin{eqnarray} 
\Big({ dE \over dt}\Big) = \big(1 + 1.8 {n_{He} \over n_{H}}\big)
m_p  n_H v
{Z^2(eff) \over A} \Big({dE \over dx}\Big)_{_{p,H}}~. 
\end{eqnarray} 
The dashed curve in Figure~14 shows this $Y(E)$ for O ions with 
$E_0$=30 MeV/nucleon, normalized such that the instantaneous 
energy deposition rate, $16\int_{0}^\infty Y(E) (dE/dt) dE$, is 
10$^{38}$ erg/s for an average ambient hydrogen density of 10 
cm$^{-3}$. This is the approximate energy deposition rate that 
accompanies the gamma-ray production rate in Orion due to O 
ions (Ramaty et al. 1996). While the energy deposition rate can 
be calculated independent of the ambient density, the 
equilibrium number does depend on the density which is quite 
uncertain. The value of 10 cm$^{-3}$ is not unreasonable if the 
accelerated particles spend part of their life time in the 
clouds. 

Next we assume that the accelerated particles are injected as a 
delta function in time (i.e. over a time interval which is very 
short compared to the life time against energy losses). The 
time-dependent differential particle number is then given by (e.g. 
Parizot et al. 1997b)
\begin{eqnarray} 
Y(E,t) = {dE'/dt \over dE/dt}  Q(E')~. 
\end{eqnarray} 
Here
\begin{eqnarray}
t = \int_E^{E'} \Big[ {dE \over dt}(E'') \Big]^{-1} dE''
\end{eqnarray}
is the elapsed time since the injection of the particles and 
the energy loss rates in equation (12) are derived at $E'$ and 
$E$. We normalize the injection source $Q(E)$ such that the 
total energy deposited at $t$=0 in the time-dependent case 
equals the deposited steady state energy over a time period of 
10$^5$ years. This leads to a total deposited energy in O 
nuclei of 3.2$\times$10$^{50}$ ergs. This value, together with 
the contributions of the other accelerated nuclei, yields a 
total accelerated particle energy content of about 10$^{51}$ 
ergs (depending on the assumed composition), which could have 
been supplied by the supernova that is thought to have occurred 
about 80,000 years ago and reheated the Orion-Eridanus bubble 
(Burrows et al. 1993).

The results are shown by the solid curves in Figure~14 for the same 
ambient density and $E_0$ as used for the steady state result. We 
see that, as time progresses, the differential number densities at 
low energies are suppressed relative to those at higher energies. 
Since in a neutral medium the K X-ray line emission from O is 
typically produced around 1 MeV/nucleon while the gamma rays are 
produced above 10 MeV/nucleon, we expect a significant reduction of 
the X-ray line to gamma-ray line production ratio relative to the 
corresponding production ratio for the steady state case. At 
t=3$\times$10$^4$ years, for example, the reduction is about an 
order of magnitude. 

\section{Conclusions}

We have investigated all the processes that lead to X-ray production 
by low energy cosmic rays for a variety of accelerated particle 
compositions and energy spectra. We demonstrated that the dominant 
continuum producing process is inverse bremsstrahlung produced by 
fast ions interacting with ambient electrons. In addition, there is 
also a significant contribution from the bremsstrahlung produced by 
secondary knock-on electrons. However, below a few keV the total 
X-ray emission produced by accelerated ions is dominated by 
relatively broad line emission (line widths $\delta E/E$$\simeq$0.1) 
resulting from de-excitations in the fast ions following electron 
captures and excitations. In addition, accelerated particle 
interactions also produce much narrower X-ray lines, due to 
inner-shell vacancy creation. The most prominent of such line is 
that at 6.4 keV from ambient Fe. 

We have calculated the X-ray line and continuum emission produced by 
the accelerated particles in Orion which are thought to be 
responsible for the nuclear gamma-ray line emission observed with 
COMPTEL (Bloemen et al. 1994, 1997). By first comparing the results 
with the extragalactic diffuse X-ray background, we found that while 
the continuum is generally below this background, the line emission 
from about 0.5 to 1.5 keV exceeds the background for all the 
combination of parameters that we considered. We wish to point out 
that there could be a significant contribution to the $\sim$0.5-1.5 
keV diffuse X-ray background from as-yet unknown sources within our 
Galaxy (e.g. Park et al. 1997), leaving the possibility that a 
substantial fraction of the observed X-ray intensity in this energy 
range results from low energy cosmic ray interactions. 

We have also compared our results with ROSAT observations of Orion 
in the 0.47 to 1.2 keV energy band, again normalizing the X-ray 
emission to the observed gamma-ray emission. We found that there is 
no conflict between the predicted total X-ray emission (lines and 
continuum) and the data for a broad range of parameters if the 
gamma-ray line emission is uniformly distributed over the entire 
molecular cloud complex. This conclusion differs from our previous 
one (Ramaty et al. 1997a) because of a lower predicted X-ray line 
emission, resulting from improved atomic physics input, and because 
our estimated ROSAT flux from Orion is higher than the upper limit 
given by Dogiel et al. (1997). However, the COMPTEL data show 
significant spatial structure. We found that for the most prominent 
hot spot in the COMPTEL map, the standard thick target, steady state 
interaction model, with a neutral ambient medium, predicts X-ray 
fluxes which exceed the ROSAT data for a broad range of parameters. 
But the calculations could be consistent with the data for any one, 
or a combination of the following possibilities: a very hard 
accelerated particle spectrum; a partially ionized ambient medium; 
and a time-dependent accelerated particle energy spectrum resulting 
from  essentially instantaneous acceleration some tens of thousand 
of years ago.

There are as-yet no astrophysical X-ray observations that would 
unambiguously indicate an origin resulting from low energy, 
accelerated ion interactions. Our calculations show that the most 
promising signatures are the relatively broad lines between 0.5 and 
1.5 keV, mainly the lines from fast O, and that a promising target 
is the Orion region where the presence of such accelerated particles 
is known from gamma-ray line observations. 

We acknowledge K. Omidvar for discussions on the atomic processes 
leading to X-ray line production. V. T. acknowledges an 
NRC-NASA/GSFC Research Associateship.

\clearpage

\begin{deluxetable}{c|c|c|c}
%\footnotesize
\tablecaption{Multiplicity of O{\sml \ VIII} line X-rays}
\tablewidth{22pc}
\tablehead{
\colhead{Elements} & \colhead{Electron capture} & 
\colhead{Excitation} & \colhead{{Totals}} \\ }
\startdata
H & 1.9 & 8.1 & 10.0 \nl
He & 4.0 & 2.2 & 6.2 \nl
C & 0.43 & 7.1$\cdot$10$^{-2}$ & 0.50 \nl
N & 0.15 & 3.1$\cdot$10$^{-2}$ & 0.18 \nl
O & 1.2 & 0.31 & 1.5 \nl
Ne & 0.23 & 7.3$\cdot$10$^{-2}$ & 0.30 \nl
Mg & 0.11 & 3.4$\cdot$10$^{-2}$ & 0.14 \nl
Si & 0.14 & 4.3$\cdot$10$^{-2}$ & 0.18 \nl
S & 8.8$\cdot$10$^{-2}$ & 3.0$\cdot$10$^{-2}$ & 0.12 \nl
Fe & 0.21 & 0.14 & 0.35 \nl
\hline
Totals & 8.5 & 11.0 & 19.5 \nl
\enddata
\end{deluxetable}

\begin{deluxetable}{c|c|c|c}
%\footnotesize
\tablecaption{Multiplicity of O{\sml \ VII} line X-rays}
\tablewidth{22pc}
\tablehead{
\colhead{Elements} & \colhead{Electron capture} & 
\colhead{Excitation} & \colhead{{Totals}} \\ }
\startdata
H & 4.1 & 12.4 & 16.5 \nl
He & 6.8 & 3.4 & 10.3 \nl
C & 0.24 & 0.11 & 0.34 \nl
N & 7.2$\cdot$10$^{-2}$ & 4.6$\cdot$10$^{-2}$ & 0.12 \nl
O & 0.59 & 0.47 & 1.1 \nl
Ne & 0.14 & 0.11 & 0.25 \nl
Mg & 6.7$\cdot$10$^{-2}$ & 5.0$\cdot$10$^{-2}$ & 0.12 \nl
Si & 7.6$\cdot$10$^{-2}$ & 6.4$\cdot$10$^{-2}$ & 0.14 \nl
S & 4.1$\cdot$10$^{-2}$ & 4.4$\cdot$10$^{-2}$ & 8.5$\cdot$10$^{-2}$ \nl
Fe & 0.10 & 0.21 & 0.32 \nl
\hline
Totals & 12.2 & 16.9 & 29.2 \nl
\enddata
\end{deluxetable}

\begin{deluxetable}{c|cc|cc|cc|cc}
%\footnotesize
\tiny
\tablecaption{Multiplicity of C, N, O and Ne K$\alpha$ lines}
\tablewidth{36pc}
\tablehead{
\colhead{$\Delta$E} & \multicolumn{2}{c}{C} & \multicolumn{2}{c}{N} & 
\multicolumn{2}{c}{O} & \multicolumn{2}{c}{Ne} \\
(MeV/nucleon) & H-like & He-like & H-like & He-like & 
H-like & He-like & H-like & He-like \\}
\startdata
 0.1 -  0.2 & 0.25E-03 & 0.42E+00 &     -    & 0.23E-01 &     -    & 0.13E-02 &     -    &     -    \nl
 0.2 -  0.3 & 0.19E-01 & 0.37E+01 & 0.54E-03 & 0.45E+00 &     -    & 0.44E-01 &     -    & 0.52E-03 \nl
 0.3 -  0.4 & 0.21E+00 & 0.90E+01 & 0.12E-01 & 0.21E+01 & 0.56E-03 & 0.35E+00 &     -    & 0.73E-02 \nl
 0.4 -  0.5 & 0.92E+00 & 0.12E+02 & 0.87E-01 & 0.45E+01 & 0.67E-02 & 0.12E+01 &     -    & 0.47E-01 \nl
 0.5 -  0.6 & 0.22E+01 & 0.11E+02 & 0.33E+00 & 0.62E+01 & 0.37E-01 & 0.23E+01 & 0.33E-03 & 0.17E+00 \nl
 0.6 -  0.7 & 0.35E+01 & 0.79E+01 & 0.80E+00 & 0.67E+01 & 0.12E+00 & 0.33E+01 & 0.20E-02 & 0.42E+00 \nl
 0.7 -  0.8 & 0.42E+01 & 0.48E+01 & 0.14E+01 & 0.60E+01 & 0.30E+00 & 0.39E+01 & 0.77E-02 & 0.77E+00 \nl
 0.8 -  0.9 & 0.41E+01 & 0.26E+01 & 0.20E+01 & 0.46E+01 & 0.56E+00 & 0.40E+01 & 0.22E-01 & 0.11E+01 \nl
 0.9 -  1.0 & 0.35E+01 & 0.13E+01 & 0.24E+01 & 0.31E+01 & 0.88E+00 & 0.36E+01 & 0.51E-01 & 0.15E+01 \nl
 1.0 -  1.5 & 0.88E+01 & 0.13E+01 & 0.96E+01 & 0.47E+01 & 0.68E+01 & 0.87E+01 & 0.13E+01 & 0.88E+01 \nl
 1.5 -  2.0 & 0.26E+01 & 0.87E-01 & 0.40E+01 & 0.41E+00 & 0.48E+01 & 0.14E+01 & 0.33E+01 & 0.50E+01 \nl
 2.0 -  2.5 & 0.10E+01 & 0.12E-01 & 0.17E+01 & 0.62E-01 & 0.24E+01 & 0.23E+00 & 0.31E+01 & 0.16E+01 \nl
 2.5 -  3.0 & 0.50E+00 & 0.29E-02 & 0.84E+00 & 0.15E-01 & 0.12E+01 & 0.56E-01 & 0.21E+01 & 0.49E+00 \nl
 3.0 -  3.5 & 0.28E+00 & 0.89E-03 & 0.47E+00 & 0.46E-02 & 0.72E+00 & 0.18E-01 & 0.14E+01 & 0.17E+00 \nl
 3.5 -  4.0 & 0.17E+00 & 0.33E-03 & 0.29E+00 & 0.17E-02 & 0.45E+00 & 0.68E-02 & 0.90E+00 & 0.65E-01 \nl
 4.0 -  4.5 & 0.11E+00 & 0.14E-03 & 0.19E+00 & 0.75E-03 & 0.30E+00 & 0.30E-02 & 0.62E+00 & 0.29E-01 \nl
 4.5 -  5.0 & 0.74E-01 &     -    & 0.13E+00 & 0.35E-03 & 0.21E+00 & 0.14E-02 & 0.44E+00 & 0.14E-01 \nl
 5.0 -  6.0 & 0.90E-01 &     -    & 0.16E+00 & 0.28E-03 & 0.26E+00 & 0.11E-02 & 0.56E+00 & 0.11E-01 \nl
 6.0 -  7.0 & 0.49E-01 &     -    & 0.92E-01 &     -    & 0.15E+00 & 0.37E-03 & 0.33E+00 & 0.37E-02 \nl
 7.0 -  8.0 & 0.29E-01 &     -    & 0.55E-01 &     -    & 0.91E-01 & 0.14E-03 & 0.21E+00 & 0.14E-02 \nl
 8.0 -  9.0 & 0.18E-01 &     -    & 0.35E-01 &     -    & 0.59E-01 &     -    & 0.14E+00 & 0.62E-03 \nl
 9.0 - 10.0 & 0.12E-01 &     -    & 0.23E-01 &     -    & 0.40E-01 &     -    & 0.94E-01 & 0.29E-03 \nl
10.0 - 12.0 & 0.13E-01 &     -    & 0.23E-01 &     -    & 0.41E-01 &     -    & 0.10E+00 & 0.16E-03 \nl
12.0 - 14.0 & 0.67E-02 &     -    & 0.13E-01 &     -    & 0.23E-01 &     -    & 0.59E-01 &     -    \nl
14.0 - 16.0 & 0.37E-02 &     -    & 0.78E-02 &     -    & 0.14E-01 &     -    & 0.37E-01 &     -    \nl
16.0 - 18.0 & 0.23E-02 &     -    & 0.50E-02 &     -    & 0.93E-02 &     -    & 0.25E-01 &     -    \nl
18.0 - 20.0 & 0.14E-02 &     -    & 0.33E-02 &     -    & 0.63E-02 &     -    & 0.17E-01 &     -    \nl
\hline
   TOTALS   &   32.53  &   54.14  &   24.66  &   38.84  &   19.55  &   29.16  &   14.79  &   20.22  \nl
\enddata
\end{deluxetable}
\normalsize

\begin{deluxetable}{c|cc|cc|cc|cc}
%\footnotesize
\tiny
\tablecaption{Multiplicity of Mg, Si, S and Fe K$\alpha$ lines}
\tablewidth{36pc}
\tablehead{
\colhead{$\Delta$E} & \multicolumn{2}{c}{Mg} & \multicolumn{2}{c}{Si} & 
\multicolumn{2}{c}{S} & \multicolumn{2}{c}{Fe} \\
(MeV/nucleon) & H-like & He-like & H-like & He-like & 
H-like & He-like & H-like & He-like \\}
\startdata
 0.5 -  1.0 & 0.21E-02 & 0.31E-01 &     -    & 0.38E-01 &     -    & 0.34E-02 &     -    &     -    \nl
 1.0 -  1.5 & 0.11E+00 & 0.12E+01 & 0.58E-02 & 0.62E+00 & 0.30E-03 & 0.11E+00 &     -    &     -    \nl
 1.5 -  2.0 & 0.66E+00 & 0.37E+01 & 0.73E-01 & 0.17E+01 & 0.69E-02 & 0.52E+00 &     -    & 0.22E-03 \nl
 2.0 -  2.5 & 0.14E+01 & 0.39E+01 & 0.30E+00 & 0.23E+01 & 0.44E-01 & 0.11E+01 &     -    & 0.17E-02 \nl
 2.5 -  3.0 & 0.17E+01 & 0.24E+01 & 0.64E+00 & 0.20E+01 & 0.14E+00 & 0.14E+01 &     -    & 0.72E-02 \nl
 3.0 -  3.5 & 0.15E+01 & 0.11E+01 & 0.90E+00 & 0.14E+01 & 0.29E+00 & 0.14E+01 &     -    & 0.20E-01 \nl
 3.5 -  4.0 & 0.12E+01 & 0.50E+00 & 0.97E+00 & 0.85E+00 & 0.45E+00 & 0.12E+01 & 0.22E-03 & 0.42E-01 \nl
 4.0 -  4.5 & 0.86E+00 & 0.23E+00 & 0.89E+00 & 0.48E+00 & 0.56E+00 & 0.86E+00 & 0.68E-03 & 0.72E-01 \nl
 4.5 -  5.0 & 0.64E+00 & 0.11E+00 & 0.76E+00 & 0.26E+00 & 0.60E+00 & 0.59E+00 & 0.17E-02 & 0.11E+00 \nl
 5.0 -  6.0 & 0.85E+00 & 0.84E-01 & 0.11E+01 & 0.23E+00 & 0.11E+01 & 0.62E+00 & 0.11E-01 & 0.32E+00 \nl
 6.0 -  7.0 & 0.51E+00 & 0.26E-01 & 0.72E+00 & 0.80E-01 & 0.85E+00 & 0.25E+00 & 0.30E-01 & 0.44E+00 \nl
 7.0 -  8.0 & 0.32E+00 & 0.97E-02 & 0.47E+00 & 0.31E-01 & 0.61E+00 & 0.10E+00 & 0.65E-01 & 0.51E+00 \nl
 8.0 -  9.0 & 0.22E+00 & 0.41E-02 & 0.32E+00 & 0.13E-01 & 0.44E+00 & 0.46E-01 & 0.11E+00 & 0.51E+00 \nl
 9.0 - 10.0 & 0.15E+00 & 0.19E-02 & 0.23E+00 & 0.64E-02 & 0.31E+00 & 0.22E-01 & 0.16E+00 & 0.45E+00 \nl
10.0 - 12.0 & 0.19E+00 & 0.15E-02 & 0.29E+00 & 0.50E-02 & 0.40E+00 & 0.17E-01 & 0.42E+00 & 0.65E+00 \nl
12.0 - 14.0 & 0.11E+00 & 0.46E-03 & 0.17E+00 & 0.16E-02 & 0.23E+00 & 0.54E-02 & 0.45E+00 & 0.34E+00 \nl
14.0 - 16.0 & 0.66E-01 & 0.17E-03 & 0.10E+00 & 0.61E-03 & 0.15E+00 & 0.20E-02 & 0.39E+00 & 0.16E+00 \nl
16.0 - 18.0 & 0.43E-01 &     -    & 0.67E-01 & 0.26E-03 & 0.97E-01 & 0.88E-03 & 0.30E+00 & 0.76E-01 \nl
18.0 - 20.0 & 0.29E-01 &     -    & 0.46E-01 & 0.12E-03 & 0.67E-01 & 0.41E-03 & 0.23E+00 & 0.37E-01 \nl
20.0 - 25.0 & 0.42E-01 &     -    & 0.67E-01 & 0.10E-03 & 0.98E-01 & 0.36E-03 & 0.36E+00 & 0.32E-01 \nl
25.0 - 30.0 & 0.21E-01 &     -    & 0.33E-01 &     -    & 0.50E-01 &     -    & 0.19E+00 & 0.79E-02 \nl
30.0 - 35.0 & 0.12E-01 &     -    & 0.18E-01 &     -    & 0.28E-01 &     -    & 0.11E+00 & 0.25E-02 \nl
35.0 - 40.0 & 0.73E-02 &     -    & 0.11E-01 &     -    & 0.17E-01 &     -    & 0.71E-01 & 0.96E-03 \nl
40.0 - 45.0 & 0.48E-02 &     -    & 0.69E-02 &     -    & 0.11E-01 &     -    & 0.48E-01 & 0.42E-03 \nl
45.0 - 50.0 & 0.33E-02 &     -    & 0.46E-02 &     -    & 0.74E-02 &     -    & 0.33E-01 & 0.20E-03 \nl
50.0 - 55.0 & 0.23E-02 &     -    & 0.32E-02 &     -    & 0.52E-02 &     -    & 0.24E-01 & 0.11E-03 \nl
55.0 - 60.0 & 0.17E-02 &     -    & 0.23E-02 &     -    & 0.38E-02 &     -    & 0.18E-01 &     -    \nl
60.0 - 65.0 & 0.13E-02 &     -    & 0.17E-02 &     -    & 0.28E-02 &     -    & 0.14E-01 &     -    \nl
65.0 - 70.0 & 0.98E-03 &     -    & 0.13E-02 &     -    & 0.21E-02 &     -    & 0.11E-01 &     -    \nl
\hline
   TOTALS   &   10.62  &   13.31  &    8.20  &   10.13  &    6.59  &    8.10  &    3.05  &    3.79  \nl
\enddata
\end{deluxetable}
\normalsize

\begin{deluxetable}{c|cc|cc}
%\footnotesize
\tablecaption{Energies in keV of K X-ray lines from fast ions. 
From Kelly (1987)}
\tablewidth{25pc}
\tablehead{
\colhead{Projectiles} & \multicolumn{2}{c}{H-like} &
\multicolumn{2}{c}{He-like} \\ & K$\alpha$ line & K$\beta$ line 
& K$\alpha$ line & K$\beta$ line \\ }
\startdata
C & 0.37 & 0.44 & 0.31 & 0.35 \nl
N & 0.50 & 0.59 & 0.43 & 0.50 \nl
O & 0.65 & 0.77 & 0.57 & 0.67 \nl
Ne & 1.02 & 1.21 & 0.92 & 1.07 \nl
Mg & 1.47 & 1.74 & 1.35 & 1.58 \nl
Si & 2.01 & 2.38 & 1.86 & 2.18 \nl
S & 2.62 & 3.11 & 2.46 & 2.88 \nl
Fe & 6.97 & 8.25 & 6.70 & 7.88 \nl
\enddata
\end{deluxetable}

\begin{deluxetable}{c|c|c}
%\footnotesize
\tablecaption{Energies in keV of K X-ray lines from ambient ions.}
\tablewidth{13pc}
\tablehead{
\colhead{Elements} & \colhead{E$_{K\alpha}$} & 
\colhead{E$_{K\beta}$} \\ }
\startdata
C  & 0.29 &  -   \nl
N  & 0.40 &  -   \nl
O  & 0.53 &  -   \nl
Ne & 0.85 &  -   \nl
Mg & 1.25 &  -   \nl
Si & 1.74 & 1.84 \nl
S  & 2.31 & 2.47 \nl
Ar & 2.95 & 3.19 \nl
S  & 3.69 & 4.01 \nl
Fe & 6.40 & 7.06 \nl
\enddata
\end{deluxetable}

\clearpage

\begin{figure}[t]   
\begin{center}     
\leavevmode 
\epsfxsize=12.cm 
\epsfbox{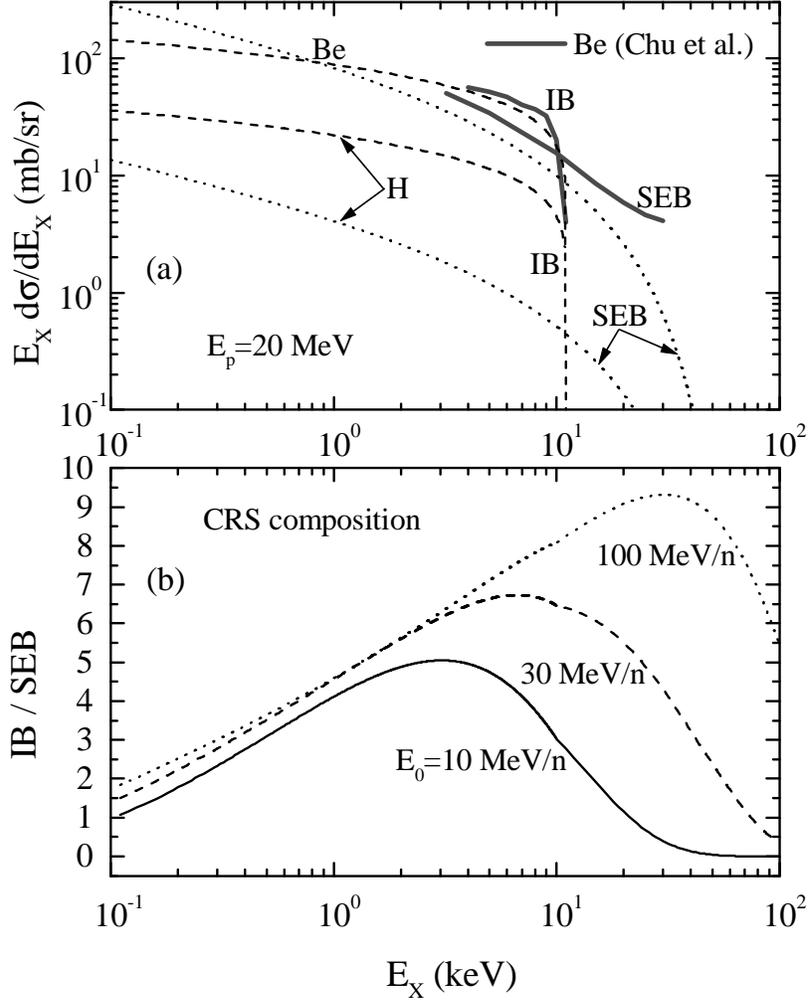}   
\end{center}   
\caption{(a): Continuum X-ray production cross sections by a 20 
MeV proton beam. The Chu et al. (1981) data (solid curves) are for 
X-rays observed from a Be target at 90$^\circ$ to the beam; the 
calculations, for both Be and H targets, are angle averaged; IB 
- inverse bremsstrahlung; SEB - secondary electron 
bremsstrahlung. (b): Ratio of the inverse bremsstrahlung (IB) 
to the secondary electron bremsstrahlung (SEB) productions, 
calculated from equation (1), for accelerated particle with CRS 
composition and spectra given by equation (2) and interacting 
with an ambient medium of solar composition.} 
\label{fig:1} 
\end{figure}

\begin{figure}[t]   
\begin{center}     
\leavevmode 
\epsfxsize=17.cm 
\epsfbox{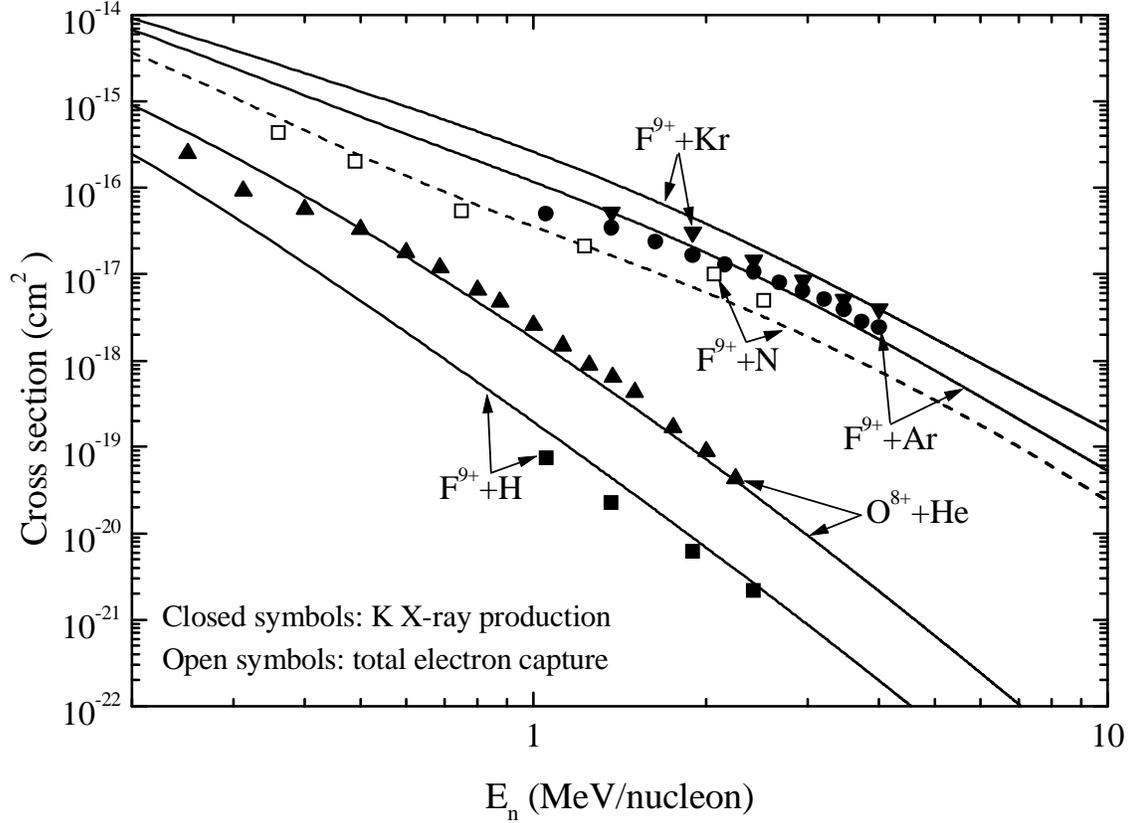}   
\end{center}   
\caption{Charge exchange cross sections. The sources of the K 
X-ray production data are: $F^{9+}+H$ -- Hopkins, Little \& 
Clue (1976a); $O^{8+}+He$ -- Guffey, Ellsworth \& Macdonald 
(1977); $F^{9+}+Ar$ and $F^{9+}+Kr$ -- Hopkins et al. (1976b). 
The $F^{9+}+N$ total electron capture data is from Macdonald et 
al. (1972). The theoretical curves employ the Nikolaev (1967) 
formalism with a single normalization constant of 0.1 for the 
five different cross sections; solid curves -- K X-ray 
production cross sections; dashed curve -- total electron 
capture cross section.} 
\label{fig:2} 
\end{figure}

\begin{figure}[t]   
\begin{center}     
\leavevmode 
\epsfxsize=17.cm 
\epsfbox{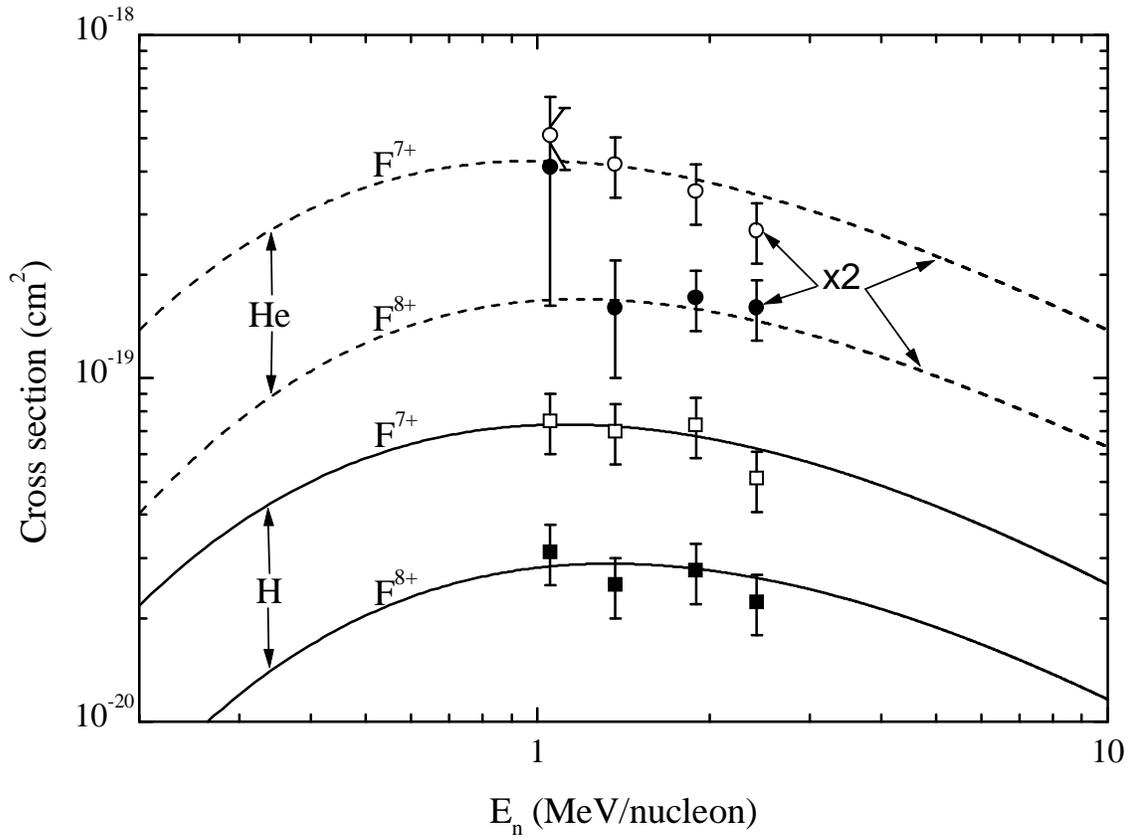}   
\end{center}   
\caption{Excitation cross sections for fast F$^{8+}$ and 
F$^{7+}$ interacting in H and He. The data are from Hopkins et 
al. (1976a). The curves were obtained by using the
plane-wave-Born-approximation (Bates 1962).} 
\label{fig:3} 
\end{figure}

\begin{figure}[t]   
\begin{center}     
\leavevmode 
\epsfxsize=12.cm 
\epsfbox{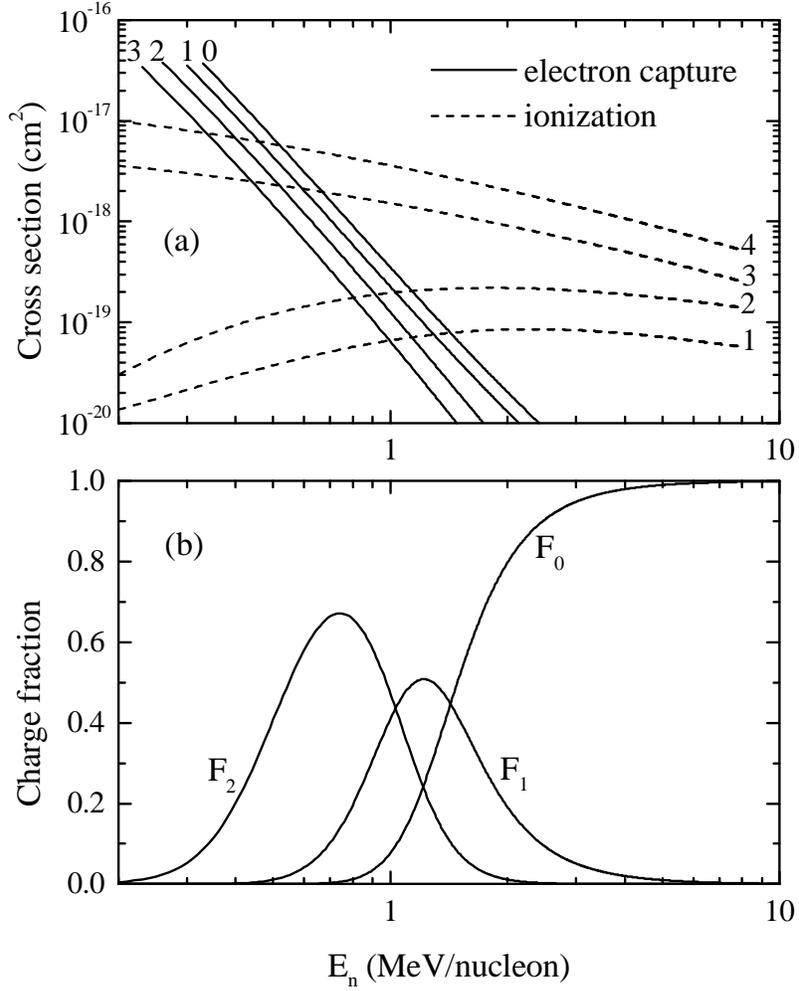}   
\end{center}   
\caption{(a): Electron capture and ionization cross sections as a 
function of kinetic energy per nucleon, of fast O interacting in an 
ambient medium of solar composition. Each curve is labeled with the 
number of bound electrons of the incident O (i.e. before the 
collision). (b): Equilibrium charge fractions as a function of 
kinetic energy of fast O in a neutral medium of solar composition.} 
\label{fig:4} 
\end{figure}

\begin{figure}[t]   
\begin{center}     
\leavevmode 
\epsfxsize=7.cm 
\epsfbox{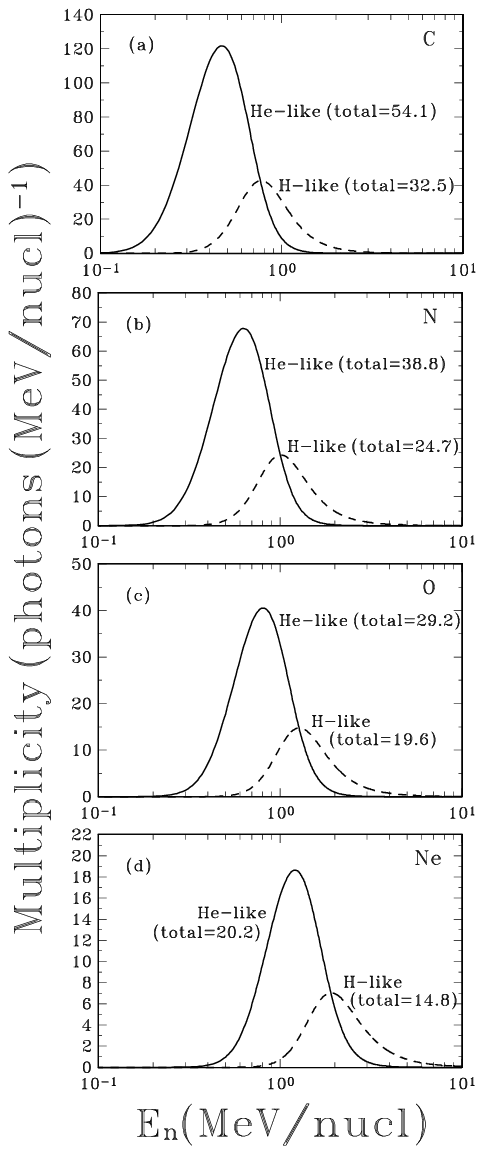}   
\end{center}   
\caption{Differential K$\alpha$ X-ray line multiplicities for 
H-like and He-like fast C, N, O and Ne slowing down in an 
ambient medium with solar abundances as functions of the  
projectile kinetic energy per nucleon $E_n$. The same 
multiplicities are listed in Table~3 and the line centroids are 
given in Table~5. The K$\beta$ multiplicities are 25\% of the 
corresponding K$\alpha$ multiplicities.} 
\label{fig:5} 
\end{figure}

\begin{figure}[t]   
\begin{center}     
\leavevmode 
\epsfxsize=7.cm 
\epsfbox{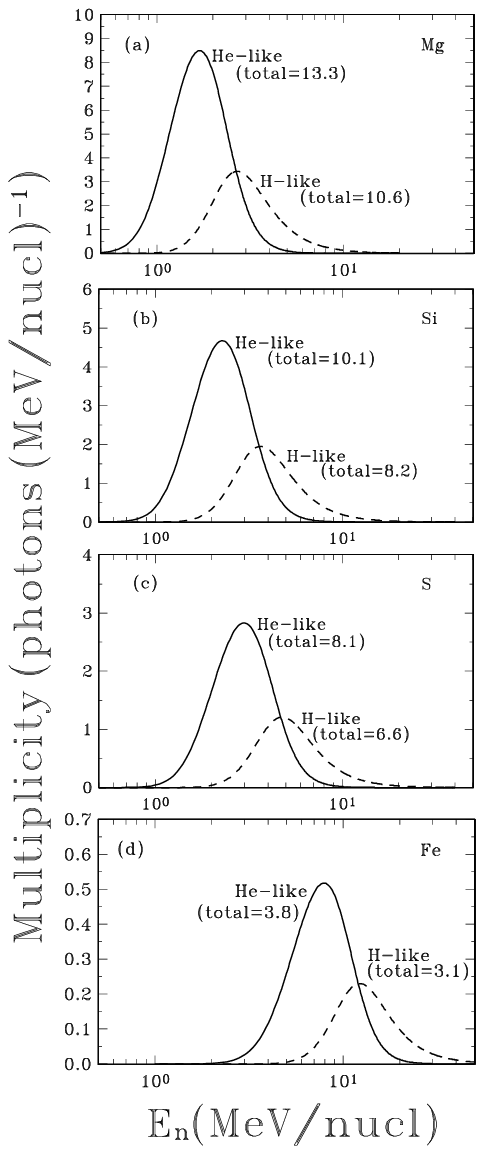}   
\end{center}   
\caption{Differential K$\alpha$ X-ray line multiplicities for 
H-like and He-like fast Mg, Si, S and Fe slowing down in an 
ambient medium with solar abundances as functions of the  
projectile kinetic energy per nucleon $E_n$. The same 
multiplicities are listed in Table~4 and the line centroids are 
given in Table~5. The K$\beta$ multiplicities are 25\% of the 
corresponding K$\alpha$ multiplicities.} 
\label{fig:6} 
\end{figure}

\begin{figure}[t]   
\begin{center}     
\leavevmode 
\epsfxsize=17.cm 
\epsfbox{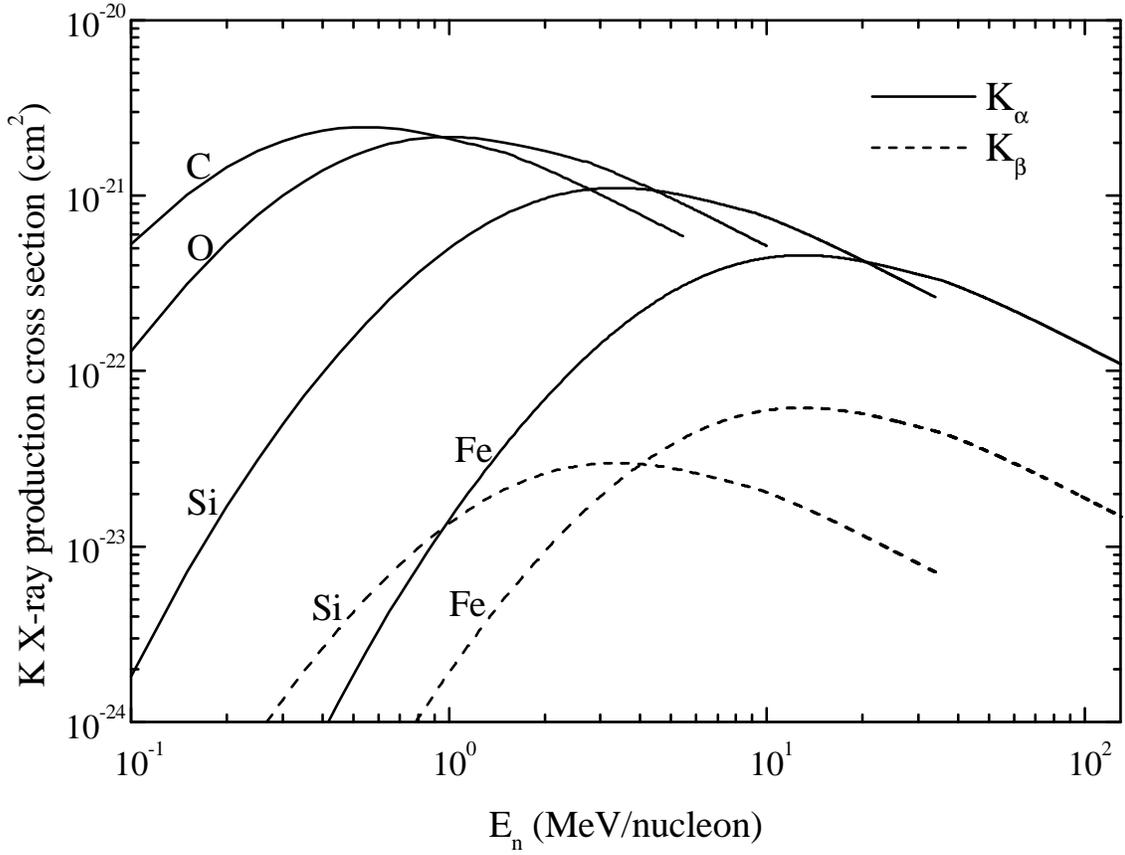}   
\end{center}   
\caption{Cross sections for the production of K X-ray lines for 
fast protons interacting with ambient neutral C, O, Si and Fe. 
We do not consider K$\beta$ X-rays from Mg and lighter atoms 
because such atoms in their ground state do not have 3$p$ 
electrons.} 
\label{fig:7} 
\end{figure}

\begin{figure}[t]   
\begin{center}     
\leavevmode 
\epsfxsize=10.cm 
\epsfbox{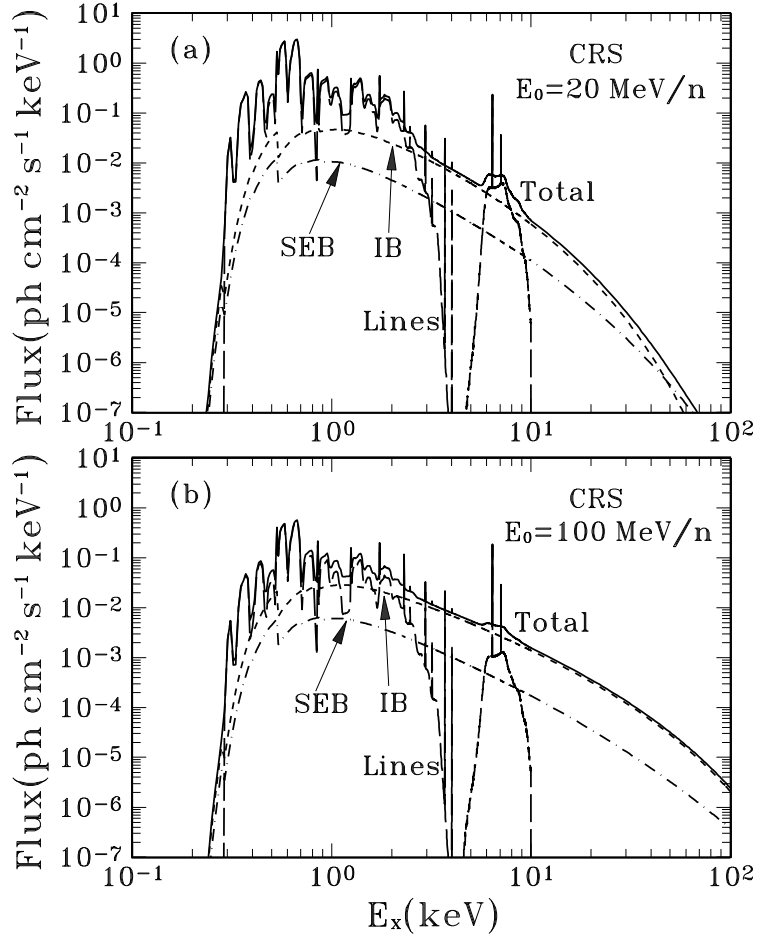}   
\end{center}   
\caption{X-ray flux for the CRS composition with two values of 
E$_0$. The calculations are normalized to a 3-7 MeV nuclear 
gamma-ray flux of 10$^{-4}$ photons cm$^{-2}$ s$^{-1}$. 
Photoelectric absorption is taken into account with a H column 
density of 3$\times$10$^{21}$ cm$^{-2}$. IB - inverse 
bremsstrahlung; SEB - secondary electron bremsstrahlung.} 
\label{fig:8} 
\end{figure}

\begin{figure}[t]   
\begin{center}     
\leavevmode 
\epsfxsize=10.cm 
\epsfbox{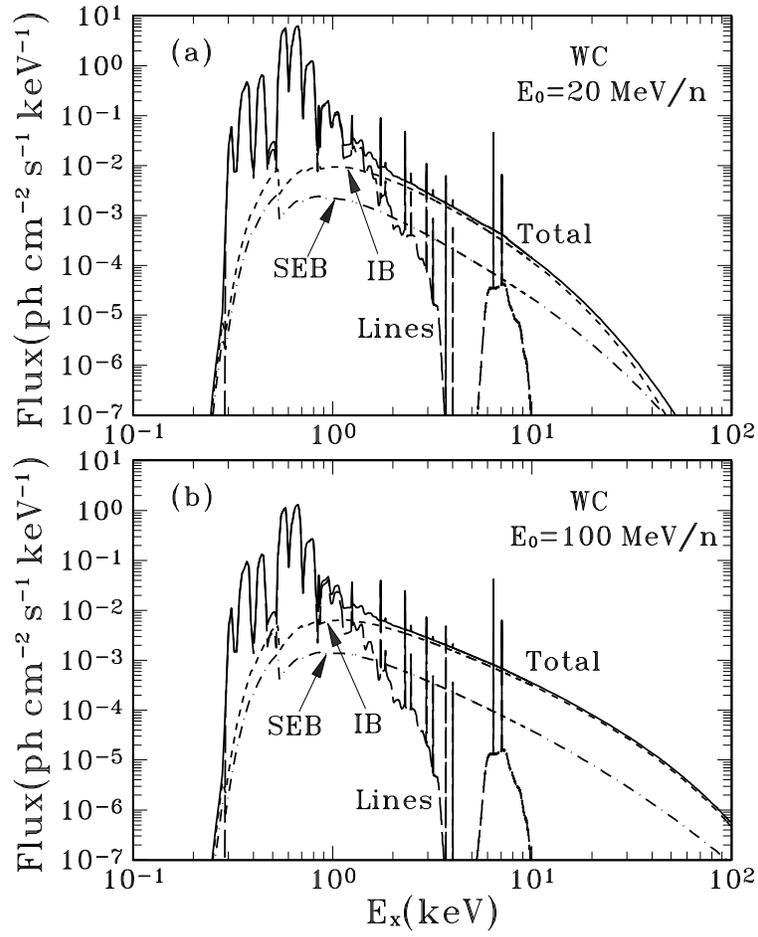}   
\end{center}   
\caption{Same as Figure~8 but for the WC composition.} 
\label{fig:9} 
\end{figure}

\begin{figure}[t]   
\begin{center}     
\leavevmode 
\epsfxsize=10.cm 
\epsfbox{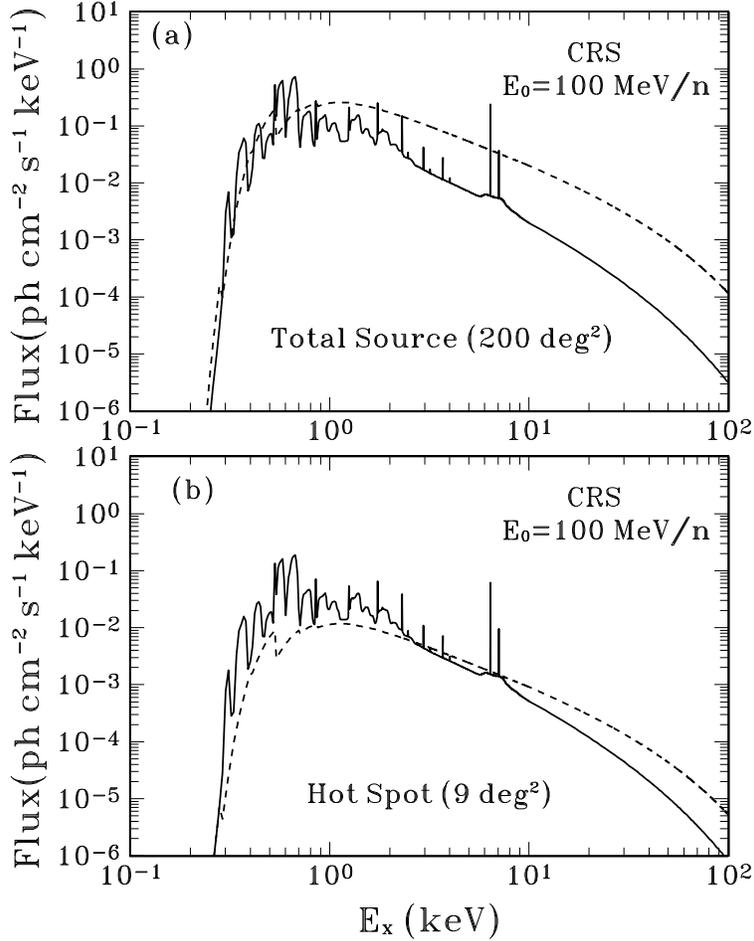}   
\end{center}   
\caption{Solid curves -- calculated X-ray fluxes from Orion 
from fast particle interactions for the CRS composition; dashed 
curves -- extragalactic X-ray background flux. Panel (a) -- 
calculated X-ray flux normalized to the total observed 3-7 MeV 
nuclear gamma-ray flux from Orion, 1.28$\times$10$^{-4}$ photons 
cm$^{-2}$ s$^{-1}$, assumed to be distributed over 200 deg$^2$; 
panel (b) -- calculated X-ray flux normalized to the observed 
nuclear gamma-ray flux of 3.3$\times$10$^{-5}$ photons 
cm$^{-2}$ s$^{-1}$ from a hot spot in Orion of assumed size 9 deg$^2$ 
centered at l$^{II}$=213.6$^{\circ}$, b$^{II}$=-15.7$^{\circ}$. 
The extragalactic flux is directly proportional to the assumed 
source size. Photoelectric absorption is taken into account 
with the same column density (N$_H$=3$\times$10$^{21}$ 
cm$^{-2}$) for the X-rays produced in Orion and the 
extragalactic background. The same column density is used for 
the sources of the two panels.} 
\label{fig:10} 
\end{figure}

\begin{figure}[t]   
\begin{center}     
\leavevmode 
\epsfxsize=10.cm 
\epsfbox{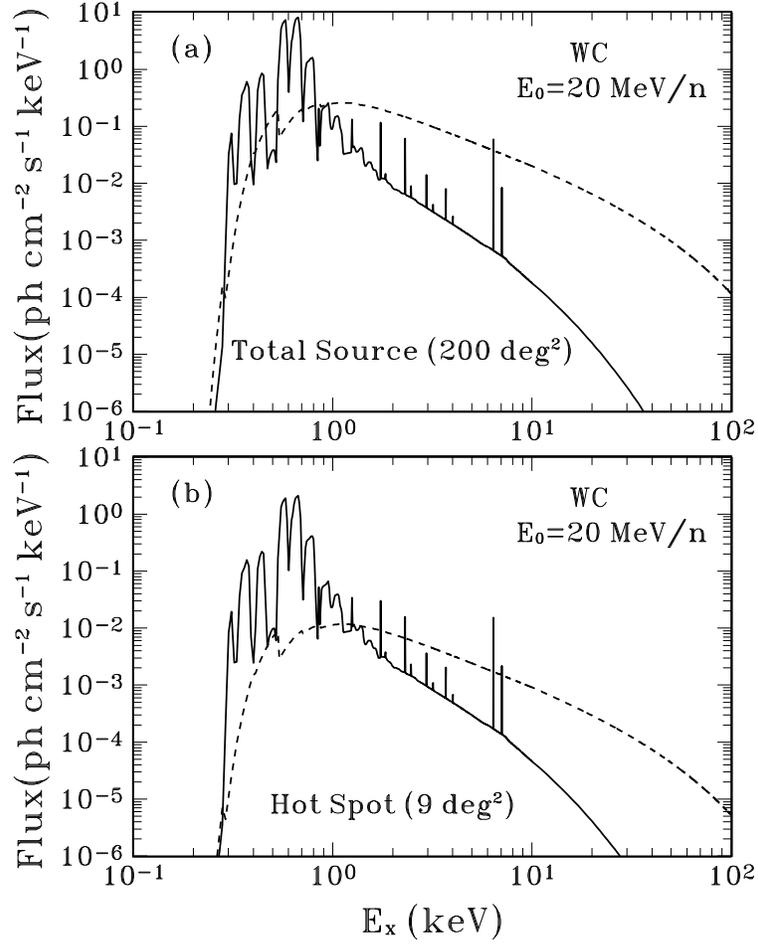}   
\end{center}   
\caption{Same as Figure~10 but for the WC composition except that different 
values of $E_0$ are used in Figures~10 and 11.} 
\label{fig:11} 
\end{figure}

\begin{figure}[t]   
\begin{center}     
\leavevmode 
\epsfxsize=12.cm 
\epsfbox{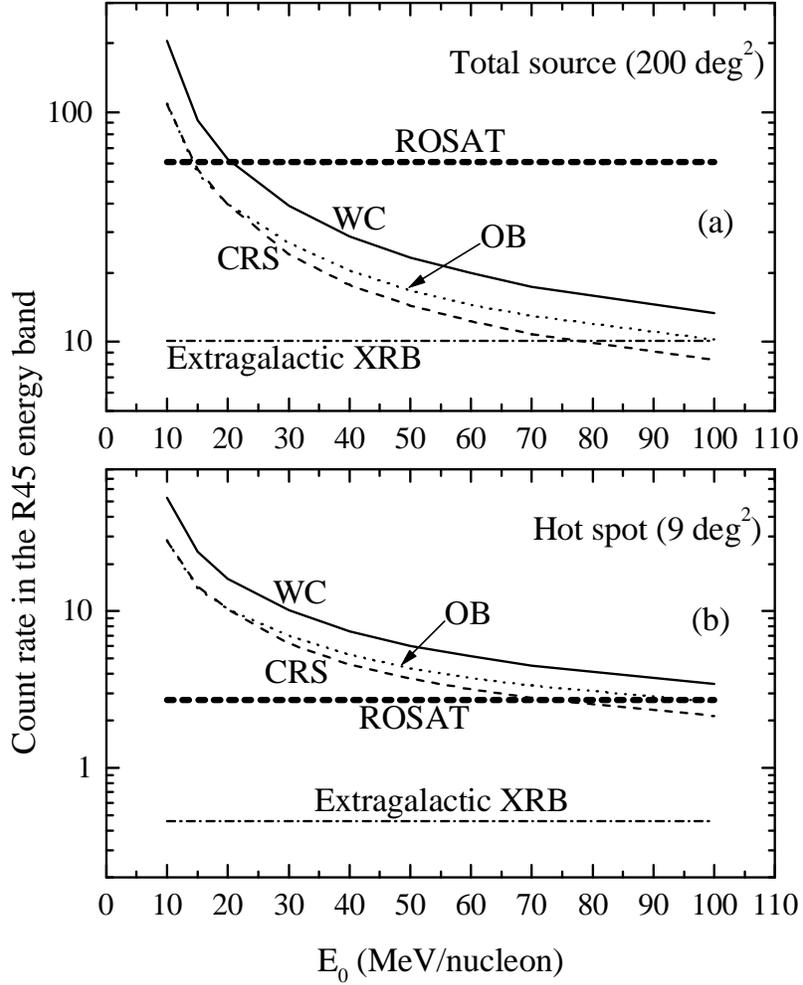}   
\end{center}   
\caption{Comparison of the ROSAT R45 data for Orion with the 
calculated X-ray emission that is expected to accompany the 
observed nuclear gamma-ray emission, as a function of the 
characteristic energy $E_0$ [Eq.~(2)] for the CRS, OB and WC 
compositions. The R45 energy band extends from 0.47 to 1.2 
keV. The calculated X-ray spectra and the extragalactic X-ray 
background spectrum are convoluted with the ROSAT/PSPC response 
function to yield the R45 count rates. Panels (a) and (b) 
correspond to the same source parameters and column density as 
panels (a) and (b) in Figures~10 and 11.} 
\label{fig:12} 
\end{figure}

\begin{figure}[t]   
\begin{center}     
\leavevmode 
\epsfxsize=12.cm 
\epsfbox{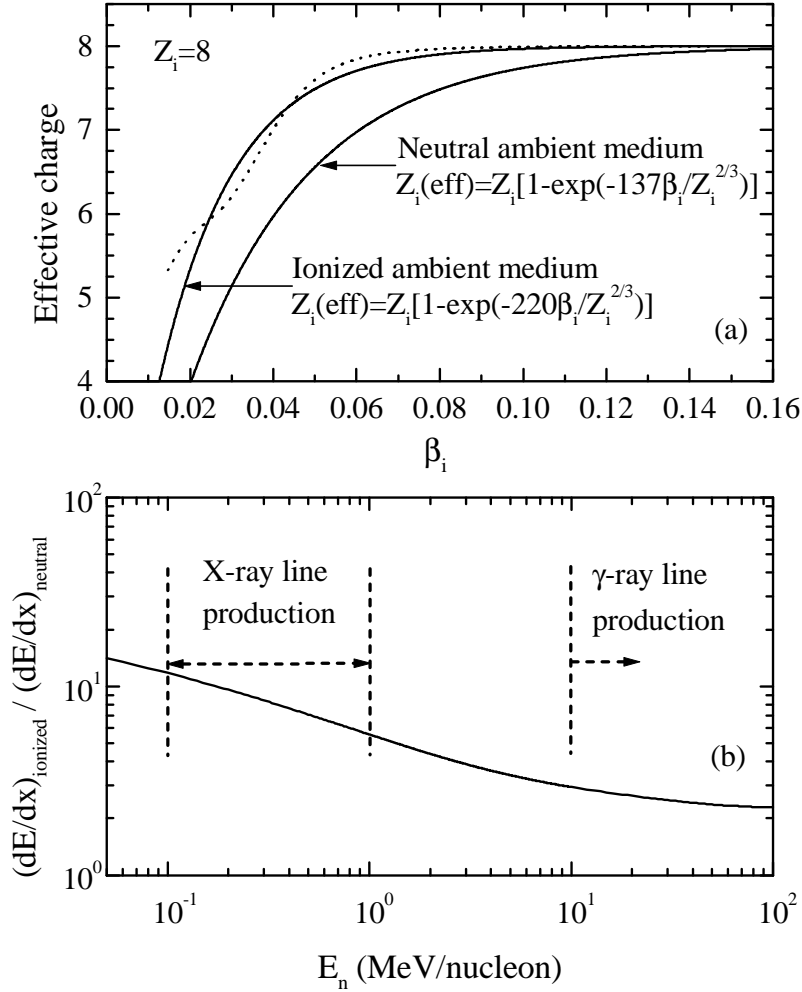}   
\end{center}   
\caption{Effective charge and energy loss of fast O in an 
ionized ambient medium of solar composition compared with those 
in a neutral medium. Panel (a): dashed curve -- calculated 
effective charge as a function of projectile velocity assuming 
that the ambient H and He are fully ionized and the heavier 
ambient elements are neutral; solid curves -- adopted fits. The 
effective charge formula for fast ions in a neutral medium is 
from Pierce and Blann (1968). Panel (b): ratio of O energy loss 
in an ionized medium to the O energy loss in a neutral medium, 
as a function of kinetic energy/nucleon. The effective charges 
are calculated as in panel (a). The temperature of the ionized 
medium is 10$^{4}$ K. Also shown are the effective energy 
ranges of X-ray line and 3-7 MeV gamma-ray line productions 
from fast O in the ionized medium.} 
\label{fig:13} 
\end{figure}

\begin{figure}[t]   
\begin{center}     
\leavevmode 
\epsfxsize=17.cm 
\epsfbox{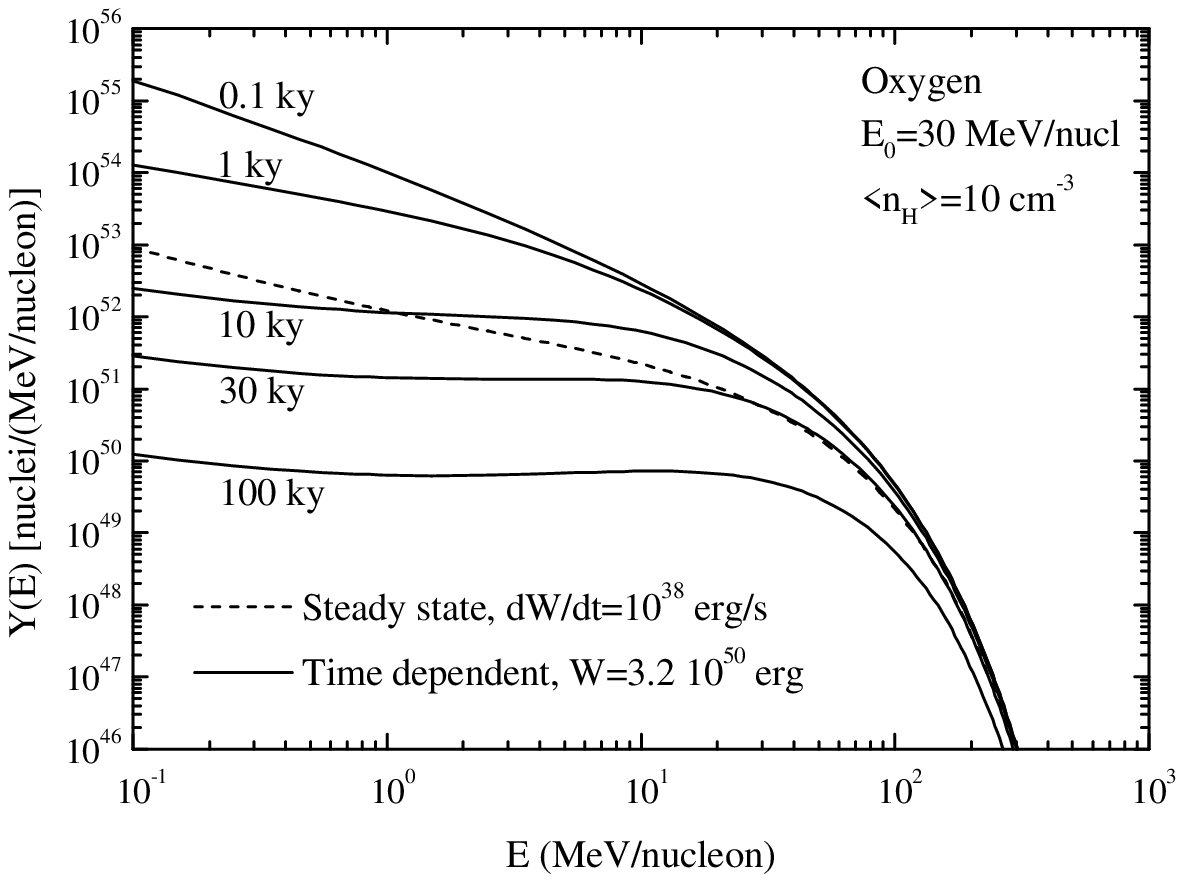}   
\end{center}   
\caption{Comparison of the instantaneous, differential accelerated 
O numbers in the steady state and the time-dependent 
models. Dashed curve -- differential equilibrium accelerated O 
number in the steady state model, normalized to a deposited 
power $dW/dt=16\int_{0}^\infty E {dN \over dt} (E) dE$ of 
10$^{38}$ erg/s. This is the approximate power required to account 
for the observed nuclear gamma-ray lines, independent of the average ambient 
density $<n_{\rm H}>$. 
Solid curves -- Time-dependent differential 
accelerated O number at different times ranging from 
0.1$\cdot$10$^3$ to 100$\cdot$10$^3$ years after the initial 
instantaneous injection, normalized to a total deposited energy 
$W=16\int_{0}^\infty E Q(E) dE$ of 3.2$\cdot$10$^{50}$ erg.
In 10$^5$ years the deposited energy for the steady state model 
will equal the total deposited energy for the time-dependent model.} 
\label{fig:14} 
\end{figure}


\clearpage

\begin{thebibliography}{}

\bibitem{} Anders, E., \& Grevesse, N. 1989, Geochim. 
Cosmochim. Acta, 53, 197

\bibitem{} Anholt, R. 1985, Rev. Mod. Phys., 57, 995

\bibitem{} Bates, D. R. 1962, Atomic and Molecular Processes 
(New York: Academic Press)

\bibitem{} Belki\'{c}, D\v{z}., Gayet, R., \& Salin, A. 1979, 
Phys. Rept., 56, 279

\bibitem{} Berger, M. J., \& Seltzer, S. M. 1982, Stopping 
Powers and Ranges of Electrons and Positrons, National Bureau 
of Standards report: NBSIR 82-2550, U.S. Dept. of Commerce, 
Washington, D.C. 20234

\bibitem{} Berkner, K. H., Graham, W. G., Pyle, R. V., 
Schlachter, A. S., \& Stearns, J. W. 1977, Phys. Lett., 62A, 407

\bibitem{} Bloemen, H., et al. 1994, A\&A, 281, L5

\bibitem{} Bloemen, H., et al. 1997, ApJ, 475, L25

\bibitem{} Boldt, E. A., \& Serlemitsos, P. 1969, ApJ, 157, 557

\bibitem{} Burrows, D. N., Singh, K. P., Nousek, J. A., 
Garmire, G. P., \& Good, J. 1993, ApJ, 406, 97

\bibitem{} Bussard, R. W., Ramaty, R., \& Omidvar, K. 1978, 
ApJ, 220, 353

\bibitem{} Bykov, A. M., \& Fleishman, G. D. 1992, MNRAS, 255, 
269

\bibitem{} Bykov, A. M., \& Bloemen, H. 1994, A\&A, 283, L1

\bibitem{} Bykov, A. M. 1995, Space Sci. Rev., 74, 397

\bibitem{} Cahill, T. A. 1980, Ann. Rev. Nucl. Part. Sci., 30, 
211

\bibitem{} Cass\'e, M., Lehoucq, R., \& Vangioni-Flam, E. 1995, 
Nature, 373, 318

\bibitem{} Chu, T. C., et al. 1981, Phys. Rev. A, 24, 1720

\bibitem{} Cowsik, R., \& Friedlander, M. 1995, ApJ, 444, L29

\bibitem{} Dame, T. M., et al. 1987, ApJ, 305, 892

\bibitem{} Digel, S. W., Hunter, S. D., \& Mukherjee, R. 1995, 
ApJ, 441, 270

\bibitem{} Dogiel, V. A., Freyberg, M. J., Morfill, G. E., \& 
Sch\"{o}nfelder, V. 1997, 25th International Cosmic Ray 
Conference, ed. M. S. Potgieter et al., vol. 3, 133

\bibitem{} Ellison, D. C., \& Ramaty, R. 1985, ApJ, 298, 400

\bibitem{} Ferguson, S. M., Macdonald, J. R., Chiao, T., 
Ellsworth, L. D., \& Savoy, S. A. 1973, Phys. Rev. A, 8, 2417

\bibitem{} Garcia, J. D., Fortner, R. J., \& Kavanagh, T. M. 
1973, Rev. Mod. Phys., 45, 111

\bibitem{} Gendreau, K. C., et al. 1995, Publ. Astron. Soc. 
Japan, 47, L5

\bibitem{} Gruber, D. E. 1992, in X-ray background, eds. X Barcons 
and A. C. Fabian (Cambridge University Press), 44

\bibitem{} Guffey, J. A., Ellsworth, L. D., \& Macdonald, J. R. 
1977, Phys. Rev. A, 15, 1863

\bibitem{} Harris, M. J., et al. 1998, A\&A, 329, 624

\bibitem{} Hayakawa, S., Matsuoka, M. 1964, Suppl. of 
the Progress of Theoretical Physics, 30, 204

\bibitem{} Hayakawa, S. 1969, Cosmic Ray Physics, Nuclear and 
Astrophysical Aspects (New York, Wiley)

\bibitem{} Hoffmann, D. H. H., et al. 1994, Nuclear Instruments 
and Methods in Physics Research B, 90, 1 

\bibitem{} Heiles, C., \& Habing, H. J. 1974, A\&AS, 14, 1

\bibitem{} Hopkins, F., Kauffman, R. L., Woods, C. W., \& 
Richard, P. 1974, Phys. Rev. A, 9, 2413

\bibitem{} Hopkins, F., Little, A., \& Cue, N. 1976a, Phys. Rev. 
A, 14, 1634

\bibitem{} Hopkins, F., et al. 1976b, Phys. Rev. A, 13, 74

\bibitem{} Ip, W.-H. 1995, A\&A, 300, 283

\bibitem{} Johansson, S. A. E., \& Johansson, T. B. 1976, Nucl. 
Inst. Meth., 137, 473

\bibitem{} Kelly, R. L. 1987, J. Phys. Chem. Ref. Data, 16, 
Suppl. 1

\bibitem{} Koch, H. W., \& Motz, J. W. 1959, 
Rev. Mod. Phys., 31, 920

\bibitem{} Kozlovsky, B., Ramaty, R., \& Lingenfelter, R. E. 
1997, ApJ, 484, 286

\bibitem{} Krause, M. O. 1979, J. Phys. Chem. Ref. Data, 8, 307

\bibitem{} Macdonald, J. R., Fergusson, S. M., Chiao, T., 
Ellsworth, L. D., \& Savoy, S. A. 1972, Phys. Rev. A, 5, 1188

\bibitem{} Mannheim, K., \& Schlickeiser, R. 1994, A\&A, 286, 983

\bibitem{} Matthews, D., Braithwaite, W. J., Wolter, H. H., \& 
Moore, C. F. 1973, Phys. Rev. A, 8, 1397

\bibitem{} Moiseiwitsch, B. L., \& Stewart, A. L. 1954, Proc. 
Phys. Soc. A, 67, 74

\bibitem{} Morrison, R., \& McCammon, D. 1983, ApJ, 270, 119

\bibitem{} Murphy, R. J., et al. 1996, ApJ, 473, 990

\bibitem{} Nath, B. B., \& Biermann, P. L. 1994, MNRAS, 270, L33

\bibitem{} Nikolaev, V. S. 1967, Soviet Phys.-JETP, 24, 847

\bibitem{} Parizot, E. M. G., Cass\'e, M., \& Vangioni-Flam, E. 
1997a, A\&A, 328, 107

\bibitem{} Parizot, E. M. G., Lehoucq, R., Cass\'e, M., \& 
Vangioni-Flam, E. 1997b, in Proc. 2nd INTEGRAL Workshop, ed. C. 
Winkler et al. (ESA SP-382), 97

\bibitem{} Park, S., Finley, J. P., Snowden, S. L., \& Dame, T. 
M. 1997, ApJ, 476, l77

\bibitem{} Pierce, T. E., \& Blann, M. 1968, Phys. Rev., 173, 
390

\bibitem{} Pravdo, S. H., \& Boldt, E. A. 1975, ApJ, 200, 727

\bibitem{} Ramaty, R. 1996, A\&AS, 120, C373

\bibitem{} Ramaty, R., Kozlovsky, B., \& Lingenfelter, R. E. 
1995, ApJ, 438, L21

\bibitem{} Ramaty, R., Kozlovsky, B., \& Lingenfelter, R. E. 
1996, ApJ, 456, 525 

\bibitem{} Ramaty, R., Kozlovsky, B., \& Tatischeff, V. 1997a, 
Proc. 4th Compton Symposium, (AIP: New York), vol. 2, 1049

\bibitem{} Ramaty, R., Kozlovsky, B., \& Lingenfelter, R. E. 
1997b, in Proc. 2nd INTEGRAL Workshop, ed. C. Winkler et al. 
(ESA SP-382), 41

\bibitem{} Raymond, J. C., \& Smith, B. W. 1977, ApJS, 
35, 419

\bibitem{} Rudd, M. E., et al. 1966, Phys. Rev., 151, 20

\bibitem{} Rule, D. W. 1977, Phys. Rev. A, 16, 19

\bibitem{} Salem, S. I., Panossian, S. L., \& Krause, R. A. 
1974, Atomic Data and Nucl. Data Tables, 14, 91

\bibitem{} Schiff, H. 1954, Canadian J. Phys., 32, 393

\bibitem{} Sevier, K. D. 1979, Atomic Data and Nucl. Data 
Tables, 24, 323

\bibitem{} Silk, J., \& Steigman, G. 1969, Phys. Rev. Lett., 
23, 597

\bibitem{} Slater, J. C. 1930, Phys. Rev., 36, 57

\bibitem{} Snowden, S. L., McCammon, D., Burrows, D. N., \& 
Mendenhall, J. A. 1994, ApJ 454, 643

\bibitem{} Snowden, S. L., et al. 1995, ApJ 454, 643

\bibitem{} Tatischeff, V., Ramaty, R., \& Mandzhavidze, N. 
1997, Proc. 4th Compton Symposium, (AIP: New York), vol.2, 1054

\bibitem{} Toburen, L. H., \& Wilson, W. E. 1972, 
Phys. Rev. A, 5, 247

\bibitem{} Watson, W. D. 1976, ApJ, 206, 842

\bibitem{} Yamauchi, S., Koyama, K., \& Inda-Koide, M. 1994, 
Publ. Astron. Soc. Japan, 46, 473

\bibitem{} Yamauchi, S., Koyama, K., Sakano, M., \& Okada, K. 
1996, Publ. Astron. Soc. Japan, 48, 719

\end{thebibliography}
\end{document}